\documentclass[12pt]{article}
\pdfoutput=1

\usepackage{booktabs}
\usepackage{tabulary}
\usepackage{multirow}
\usepackage{rotating}
\usepackage{epsfig}
\usepackage{psfrag}
\usepackage{latexsym}
\usepackage{indentfirst}
\usepackage{fancyhdr}
\usepackage{dsfont}
\usepackage{amsmath}
\usepackage{amssymb}
\usepackage{amsfonts}
\usepackage{mathrsfs}
\usepackage{amsthm}
\usepackage{pifont}
\usepackage{dsfont}
\usepackage{multirow}
\usepackage{array}
\usepackage{chngpage}
\usepackage{longtable}
\usepackage{cite}
\usepackage{bbold}
\usepackage{color}
\usepackage{braket}
\usepackage{colordvi}
\usepackage{fancybox}
\usepackage[footnotesize]{caption2}
\usepackage{graphicx}
\usepackage[center,footnotesize,hang]{subfigure}
\usepackage{bbm}
\usepackage{url}
\usepackage[colorlinks, linkcolor=red, anchorcolor=black, citecolor=green]{hyperref}
\usepackage{multirow}
\usepackage{harpoon}
\usepackage{array}
\usepackage{textcomp}  
\usepackage{enumitem}
\usepackage{mathrsfs}  
\newcommand{\PreserveBackslash}[1]{\let\temp=\\#1\let\\=\temp}
\newcolumntype{C}[1]{>{\PreserveBackslash\centering}p{#1}}
\newcolumntype{R}[1]{>{\PreserveBackslash\raggedleft}p{#1}}
\newcolumntype{L}[1]{>{\PreserveBackslash\raggedright}p{#1}}
\allowdisplaybreaks  

\newcommand{\bq}{\begin{eqnarray}}
\newcommand{\nq}{\end{eqnarray}}

\begin{document}
\title{
\begin{flushright}
\hfill\mbox{\small USTC-ICTS-19-16} \\[5mm]
\begin{minipage}{0.2\linewidth}
\normalsize
\end{minipage}
\end{flushright}
{\Large \bf
Modular $A_4$ Symmetry Models of Neutrinos and Charged Leptons
\\[2mm]}}
\date{}

\author{
Gui-Jun~Ding$^{1}$\footnote{E-mail: {\tt
dinggj@ustc.edu.cn}},  \
Stephen~F.~King$^{2}$\footnote{E-mail: {\tt king@soton.ac.uk}}, \
Xiang-Gan Liu$^{1}$\footnote{E-mail: {\tt
hepliuxg@mail.ustc.edu.cn}}  \
\\*[20pt]
\centerline{
\begin{minipage}{\linewidth}
\begin{center}
$^1${\it \small
Interdisciplinary Center for Theoretical Study and  Department of Modern Physics,\\
University of Science and Technology of China, Hefei, Anhui 230026, China}\\[2mm]
$^2${\it \small
Physics and Astronomy,
University of Southampton,
Southampton, SO17 1BJ, U.K.}\\
\end{center}
\end{minipage}}
\\[10mm]}
\maketitle
\thispagestyle{empty}

\begin{abstract}
We present a comprehensive analysis of neutrino mass and lepton mixing in theories with $A_4$ modular symmetry, where the only flavon field is the single modulus field $\tau$, and all masses and Yukawa couplings are modular forms. Similar to previous analyses, we discuss all the simplest neutrino sectors arising from both the Weinberg operator and the type I seesaw mechanism, with lepton doublets and right-handed neutrinos assumed to be triplets of $A_4$. Unlike previous analyses, we allow right-handed charged leptons to transform as all combinations of $\mathbf{1}$, $\mathbf{1}'$ and $\mathbf{1}''$ representations of $A_4$, using the simplest different modular weights to break the degeneracy, leading to ten different charged lepton Yukawa matrices, instead of the usual one. This implies ten different Weinberg models and thirty different type I seesaw models, which we analyse in detail. We find that fourteen models for both NO and IO neutrino mass ordering can accommodate the data, as compared to one in previous analyses, providing many new possibilities.
\end{abstract}
\newpage

\section{\label{sec:introduction}Introduction}

Despite the measurement of a non-zero reactor angle, it remains an intriguing possibility that
the large mixing angles in the lepton sector can be explained using some discrete non-Abelian family symmetry \cite{King:2013eh,King:2017guk}. The origin of such a symmetry could either
be a continuous non-Abelian gauge symmetry, broken to a discrete subgroup
\cite{Koide:2007sr,Banks:2010zn,Wu:2012ria,Merle:2011vy,Rachlin:2017rvm,Luhn:2011ip,King:2018fke},
or it could emerge from extra dimensions \cite{Altarelli:2008bg,Burrows:2009pi,Burrows:2010wz,deAnda:2018oik,Adulpravitchai:2009id,Asaka:2001eh,Altarelli:2006kg,Adulpravitchai:2010na,Kobayashi:2006wq,deAnda:2018yfp,Kobayashi:2018rad,Baur:2019kwi}, either as an accidental symmetry
of the orbifold fixed points, or as a subgroup of the symmetry of the extra dimensional lattice vectors, commonly referred to as modular symmetry~\cite{Giveon:1988tt,Altarelli:2005yx,deAdelhartToorop:2011re}.

Recently it has been suggested that neutrino masses might be modular forms
\cite{Feruglio:2017spp}, with constraints on the Yukawa couplings.
The idea is that, since modular invariance controls orbifold compactifications of the heterotic superstring,
this implies that the 4d effective Lagrangian must respect modular symmetry, hence the Yukawa couplings (involving twisted states whose modular weights do not add up to zero) are modular forms \cite{Feruglio:2017spp}. Hence the
Yukawa couplings form multiplets with well defined alignments, prescribed by the modular form, which depend on a single complex modulus field $\tau$.

This has led to a revival of the idea that  modular symmetries are symmetries of the extra dimensional spacetime
with Yukawa couplings determined by their modular weights \cite{Feruglio:2017spp,Criado:2018thu}.
The finite modular subgroups considered in the literature include
$\Gamma(2)$~\cite{Kobayashi:2018vbk,Kobayashi:2018wkl,Kobayashi:2019rzp,Okada:2019xqk}, $\Gamma(3)$~\cite{Feruglio:2017spp,Criado:2018thu,Kobayashi:2018vbk,Kobayashi:2018scp,Okada:2018yrn,Kobayashi:2018wkl,Novichkov:2018yse,Nomura:2019yft}, $\Gamma(4)$~\cite{Penedo:2018nmg,Novichkov:2018ovf,Kobayashi:2019mna} and $\Gamma(5)$~\cite{Novichkov:2018nkm,Ding:2019xna}.
The $\Gamma(3)$ case has been applied to grand unified theories with the modulus fixed by the orbifold construction
\cite{deAnda:2018ecu}. The formalism with a single complex modulus field $\tau$ has also been extended to the case of
multiple moduli fields $\tau_i$~\cite{deMedeirosVarzielas:2019cyj}. The generalized CP symmetry in modular invariant models are studied in~\cite{Novichkov:2019sqv}. The formalism of modular invariant approach has extended to include odd weight modular forms~\cite{Liu:2019khw}.

In this paper, we shall study the finite modular group
$\Gamma_3\cong A_4$ with a single modulus field $\tau$ and no other flavons, hence all masses and Yukawa couplings are modular forms.
Similar to previous analyses~\cite{Feruglio:2017spp,Criado:2018thu,Kobayashi:2018scp,Okada:2018yrn,Kobayashi:2018wkl,Kobayashi:2018vbk,Novichkov:2018yse,Nomura:2019yft},
we discuss all the simplest neutrino sectors arising from both the Weinberg operator and the type I seesaw mechanism, with lepton doublets and right-handed neutrinos assumed to be triplets of $A_4$.
However, unlike all previous analyses~\cite{Feruglio:2017spp,Criado:2018thu,Kobayashi:2018scp,Okada:2018yrn,Kobayashi:2018wkl,Kobayashi:2018vbk,Novichkov:2018yse,Nomura:2019yft},
we allow right-handed charged leptons to
transform as all combinations of $\mathbf{1}$, $\mathbf{1}'$ and $\mathbf{1}''$
representations of $A_4$, using the simplest different modular weights to break the degeneracy, leading to ten different charged lepton Yukawa matrices, instead of the usual one. This implies ten different Weinberg models and thirty different type I seesaw models, which we analyse in detail. We find that fourteen models for both normal ordering (NO) and inverted ordering (IO) neutrino mass spectrums can accommodate the data, as compared to one in previous analyses, providing many new possibilities.

The structure of the paper is as follows. In section~\ref{sec:rev_mods} we briefly outline the idea of modular symmetry, and we specialize to $\Gamma(3)$ modular symmetry and give the modular forms of level $N=3$. Then in section~\ref{sec:A4models} we systematically construct and classify
the forty simplest models based on $\Gamma_3\cong A_4$,
generalising previous analyses in the charged lepton sector as outlined above. After that in section~\ref{sec:numerical} we perform a comprehensive and systematic numerical analysis for each of the forty models discussed in the previous section, giving the best fit values of the parameters for each viable model with NO and the corresponding predictions in a detailed compendium of tables and figures. Section~\ref{sec:conclusion} concludes the paper.

\section{\label{sec:rev_mods}Modular symmetry and modular forms of level $N=3$}

In the following, we briefly review the modular symmetry and the its congruence subgroups. The special linear group $SL(2,\mathbb{Z})$ is constituted by $2\times2$ matrices with integer entries and
determinant 1~\cite{Bruinier2008The,diamond2005first}:
\begin{equation}
SL(2,\mathbb{Z})=\left\{\left(\begin{array}{cc}a&b\\c&d\end{array}\right)\bigg|a,b,c,d\in \mathbb{Z},ad-bc=1\right\}\,.
\end{equation}
The upper half plane, denoted as $\mathcal{H}$, is the set of all complex numbers with positive imaginary part: $\mathcal{H}=\left\{\tau\in\mathbb{C}~|~\Im\tau >0\right\}$. The $SL(2,\mathbb{Z})$ group acts on $\mathcal{H}$ via fractional linear transformations (or M$\ddot{\mathrm{o}}$bius transformations),
\begin{equation}
\label{eq:modular_trans}\gamma=\begin{pmatrix}
a  &  b  \\
c  &  d
\end{pmatrix}:\mathcal{H}\rightarrow\mathcal{H},\quad \tau \mapsto \gamma\tau=\gamma(\tau)=\frac{a\tau+b}{c\tau+d}\,.
\end{equation}
It is straightforward to check that
\begin{equation}
\Im(\gamma(\tau))=\frac{\Im\tau}{|c\tau+d|^2},\quad \gamma=\begin{pmatrix}
a  &  b  \\
c  &  d
\end{pmatrix}\in SL(2, \mathbb{Z})\,,
\end{equation}
which implies if $\gamma\in SL(2, \mathbb{Z})$ and $\tau\in\mathcal{H}$ then also $\gamma(z)\in\mathcal{H}$. Therefore the modular group maps the upper half plane back to itself. In fact the modular group acts on the upper half plane, meaning that $I(\tau)=\tau$ where $I$ is the $2\times2$ identity matrix and $(\gamma\gamma')(\tau)=\gamma(\gamma'(\tau))$ for any $\gamma, \gamma'\in SL(2, \mathbb{Z})$ and $\tau\in\mathcal{H}$. Furthermore, $\gamma$ and $-\gamma$ evidently give the same action, therefore it is more natural to consider the projective special linear group $PSL(2, \mathbb{Z})=SL(2, \mathbb{Z})/\{I, -I\}$, the quotient of $SL(2, \mathbb{Z})$ by $\pm I$. The group $PSL(2, \mathbb{Z})$ is usually called the modular group in the literature, and it can be generated by
two elements $S$ and $T$~\cite{Bruinier2008The}
\begin{equation}
S=\left(
\begin{array}{cc}
0 ~&~ 1\\
-1 ~&~ 0
\end{array}
\right),\quad  T=\left(
\begin{array}{cc}
1 ~&~  1\\
0 ~&~ 1
\end{array}
\right)\,,
\end{equation}
which satisfy the relations
\begin{equation}
S^2=(ST)^3=\mathds{1}\,.
\end{equation}
The actions of $S$ and $T$ on $\mathcal{H}$ are given by
\begin{equation}
S: \tau \mapsto -\frac{1}{\tau},\qquad T: \tau \mapsto \tau+1\,.
\end{equation}
For a positive integer $N$, the principal congruence subgroup of level $N$ of is defined as
\begin{equation}
\Gamma(N)=\left\{\left(\begin{array}{cc}a&b\\c&d\end{array}\right)\in SL(2,\mathbb{Z}),~~a\equiv d\equiv1~({\tt mod}~N),~~ b\equiv c\equiv0~({\tt mod}~N) \right\}\,,
\end{equation}
which is a normal subgroup of the special linear group $SL(2, \mathbb{Z})$.  Obviously $\Gamma(1)\cong SL(2, \mathbb{Z})$ is the special linear group. It is easy to obtain
\begin{equation}
T^{N}=\begin{pmatrix}
1  ~&~  N \\
0   ~&~  1
\end{pmatrix}\,,
\end{equation}
which implies $T^N\in\Gamma(N)$, i.e., $T^N$ is an element of $\Gamma(N)$. Taking the quotient of $\Gamma(1)$ and $\Gamma(2)$ by $\{I, -I\}$, we obtain the projective principal congruence subgroups $\overline{\Gamma}(N)=\Gamma(N)/\{I, -I\}$ for $N=1, 2$, and $\overline{\Gamma}(N>2)=\Gamma(N)$ since the element $-I$ doesn't belong to $\Gamma(N)$ for $N>2$. The quotient groups $\Gamma_{N}=\overline{\Gamma}(1)/\overline{\Gamma}(N)$ are usually called finite modular groups, and the group $\Gamma_N$ can be obtained from $\overline{\Gamma}(1)$ by imposing the condition $T^N=1$. Consequently the generators $S$ and $T$ of $\Gamma_N$ satisfy the relations
\begin{equation}
S^2=(ST)^3=T^{N}=\mathds{1}\,.
\end{equation}
The groups $\Gamma_N$ with $N=2$, $3$, $4$, $5$ are isomorphic to the permutation groups $S_3$, $A_4$, $S_4$ and $A_5$ respectively~\cite{deAdelhartToorop:2011re}.

The crucial element of modular invariance approach is the modular form $f(\tau)$ of weight $k$ and level $N$. The modular form $f(\tau)$ is a holomorphic function of the complex modulus $\tau$ and it is required to satisfy the following modular transformation property under the group $\Gamma(N)$,
\begin{equation}
f\left(\frac{a\tau+b}{c\tau+d}\right)=(c \tau+d)^{k} f(\tau) \quad\text{for}\quad \forall~~\gamma=\begin{pmatrix}
a  &  b \\
c  &  d
\end{pmatrix}\in \Gamma(N)~~\text{and}~~\tau\in\mathcal{H}\,.
\end{equation}
The modular forms of weight $k$ and level $N$ span a linear space $\mathcal{M}_k(\Gamma(N))$ with finite dimension. As has been shown in~\cite{Feruglio:2017spp,Liu:2019khw}, we can choose the basis vectors of $\mathcal{M}_k(\Gamma(N))$ such that they can be organized into multiplets of modular forms $F_\mathbf{r}(\tau)\equiv\left(f_1(\tau),\,f_2(\tau),\,\dots\right)^T$ which transform in certain irreducible representation of the finite modular group $\Gamma_N$,
\begin{equation}
\label{eq:decomp_MF}F_\mathbf{r}(\gamma\tau)= (c\tau + d)^k\rho_{\mathbf{r}}(\gamma)F_\mathbf{r}(\tau),\quad \gamma \in \overline{\Gamma}(1)\,,
\end{equation}
where $\gamma$ is the representative element of the coset $\gamma\overline{\Gamma}(N)$ in $\Gamma_N$, and $\rho_{\mathbf{r}}(\gamma)$ is the representation matrix of the element $\gamma$ in the irreducible representation $\mathbf{r}$. When $\gamma$ is the generators $S$ and $T$, Eq.~\eqref{eq:decomp_MF} gives
\begin{equation}
\label{eq:decom_ST}F_\mathbf{r}(S\tau)= \tau^k\rho_{\mathbf{r}}(S)F_\mathbf{r}(\tau),\quad F_\mathbf{r}(T\tau)= \rho_{\mathbf{r}}(T)F_\mathbf{r}(\tau)\,,
\end{equation}
for even $k$.

\subsection{Modular forms of level $N=3$}

In the present work, we present a comprehensive analysis of neutrino mass and lepton mixing in theories with $\Gamma(3)$ modular symmetry.
The finite modular group $\Gamma_3$ is isomorphic to $A_4$ which is the symmetry group of the tetrahedron. It contains twelve elements and it is the smallest non-abelian finite group which admits a three-dimensional irreducible representation. The $A_4$ group has three singlet representations $\mathbf{1}$, $\mathbf{1}'$, $\mathbf{1}''$ and a triplet representation $\mathbf{3}$. In the singlet representations, we have
\begin{equation}
\begin{aligned}
& \mathbf{1}:~~ S=1, \qquad T=1 \,,  \\
& \mathbf{1}^{\prime}:~~ S=1, \qquad T=\omega^{2} \,,  \\
&\mathbf{1}^{\prime\prime}:~~S=1, \qquad T=\omega \,,
\end{aligned}
\end{equation}
with $\omega=e^{2\pi i/3}=-1/2+i\sqrt{3}/2$. For the representation $\mathbf{3}$, we will choose a basis in which the generator $T$ is diagonal. The explicit forms of $S$ and $T$ are
\begin{equation}
S=\frac{1}{3}\begin{pmatrix}
    -1& 2  & 2  \\
    2  & -1  & 2 \\
    2 & 2 & -1
\end{pmatrix}, ~\quad~
T=\begin{pmatrix}
    1 ~&~ 0 ~&~ 0 \\
    0 ~&~ \omega^{2} ~&~ 0 \\
    0 ~&~ 0 ~&~ \omega
\end{pmatrix} \,,
\end{equation}
The basic multiplication rule is
\begin{equation}
\mathbf{3}\otimes \mathbf{3}= \mathbf{1}\oplus \mathbf{1'}\oplus \mathbf{1''}\oplus \mathbf{3}_S\oplus \mathbf{3}_A\,,
\end{equation}
where the subscripts $S$ and $A$ denotes symmetric and antisymmetric combinations respectively. If we have two triplets $\alpha=(\alpha_1,\alpha_2,\alpha_3)\sim\mathbf{3}$ and  $\beta=(\beta_1,\beta_2,\beta_3)\sim\mathbf{3}$, we can obtain the following irreducible representations from their product,
\begin{eqnarray}
\nonumber &&(\alpha\beta)_{\mathbf{1}}=\alpha_1\beta_1+\alpha_2\beta_3+\alpha_3\beta_2\,, \\
\nonumber &&(\alpha\beta)_{\mathbf{1}^{\prime}}=\alpha_3\beta_3+\alpha_1\beta_2+\alpha_2\beta_1\,, \\
\nonumber &&(\alpha\beta)_{\mathbf{1}^{\prime\prime}}=\alpha_2\beta_2+\alpha_1\beta_3+\alpha_3\beta_1\,, \\
\nonumber &&(\alpha\beta)_{\mathbf{3}_S}=(
2\alpha_1\beta_1-\alpha_2\beta_3-\alpha_3\beta_2,
2\alpha_3\beta_3-\alpha_1\beta_2-\alpha_2\beta_1,
2\alpha_2\beta_2-\alpha_1\beta_3-\alpha_3\beta_1)\,, \\
\label{eq:CG_coefficient} &&(\alpha\beta)_{\mathbf{3}_A}=(
\alpha_2\beta_3-\alpha_3\beta_2,
\alpha_1\beta_2-\alpha_2\beta_1,
\alpha_3\beta_1-\alpha_1\beta_3)\,.
\end{eqnarray}
The linear space of the modular forms of integral weight $k$ and level $N=3$  has dimension $k+1$~\cite{Gunning1962,Feruglio:2017spp}. The modular space $\mathcal{M}_{2k}(\Gamma(3))$ can be constructed from the Dedekind eta-function $\eta(\tau)$ which is defined as
\begin{equation}
\eta(\tau)=q^{1/24} \prod_{n =1}^\infty (1-q^n), \qquad  q=e^{2\pi i\tau}\,.
\end{equation}
The eta function $\eta(\tau)$ satisfies the following identities
\begin{equation}
\eta(\tau+1)=e^{i \pi/12}\eta(\tau),\qquad \eta(-1/\tau)=\sqrt{-i \tau}~\eta(\tau)\,.
\end{equation}
There are only three linearly independent modular forms of weight 2 and level 3, which are denoted as $Y_i(\tau)$ with $i=1, 2, 3$. We can arrange the three modular functions into a vector $Y^{(2)}_{\mathbf{3}}=\left(Y_1, Y_2, Y_3\right)^{T}$ transforming as a triplet $\mathbf{3}$ of $A_4$. The modular forms $Y_i$ can be expressed in terms of $\eta(\tau)$ and its derivative as follow~\cite{Feruglio:2017spp}:
\begin{eqnarray}
Y_1(\tau) &=& \frac{i}{2\pi}\left[ \frac{\eta'(\tau/3)}{\eta(\tau/3)}  +\frac{\eta'((\tau +1)/3)}{\eta((\tau+1)/3)}
+\frac{\eta'((\tau +2)/3)}{\eta((\tau+2)/3)} - \frac{27\eta'(3\tau)}{\eta(3\tau)}  \right], \nonumber \\
Y_2(\tau) &=& \frac{-i}{\pi}\left[ \frac{\eta'(\tau/3)}{\eta(\tau/3)}  +\omega^2\frac{\eta'((\tau +1)/3)}{\eta((\tau+1)/3)}
+\omega \frac{\eta'((\tau +2)/3)}{\eta((\tau+2)/3)}  \right] ,\nonumber \\
Y_3(\tau) &=& \frac{-i}{\pi}\left[ \frac{\eta'(\tau/3)}{\eta(\tau/3)}  +\omega\frac{\eta'((\tau +1)/3)}{\eta((\tau+1)/3)}
+\omega^2 \frac{\eta'((\tau +2)/3)}{\eta((\tau+2)/3)} \right]\,.
\end{eqnarray}
Notice that $12\eta'(\tau)/\eta(\tau)\equiv i\pi E_2(\tau)$, where $E_2(\tau)$ is the well-known Eisenstein series of weight 2~\cite{Bruinier2008The}. The $q$-expansions of the triplet modular forms $Y^{(2)}_{\mathbf{3}}$ are given by
\begin{eqnarray}
Y^{(2)}_{\mathbf{3}}=\begin{pmatrix}Y_1(\tau)\\Y_2(\tau)\\Y_3(\tau)\end{pmatrix}=
\begin{pmatrix}
1 + 12q + 36q^2 + 12q^3 + 84q^4 + 72q^5 +\dots \\
-6q^{1/3}(1 + 7q + 8q^2 + 18q^3 + 14q^4 +\dots) \\
-18q^{2/3}(1 + 2q + 5q^2 + 4q^3 + 8q^4 +\dots)
\end{pmatrix}\,.
\end{eqnarray}
They satisfy the constraint~\cite{Feruglio:2017spp,Liu:2019khw}
\begin{equation}
\label{eq:MF2}(Y^{(2)}_{\mathbf{3}}Y^{(2)}_{\mathbf{3}})_{\mathbf{1}''}=Y_2^2+2 Y_1 Y_3=0\,.
\end{equation}
Multiplets of higher weight modular forms can be constructed from the tensor products of $Y^{(2)}_{\mathbf{3}}$. Using the $A_4$ contraction rule $\mathbf{3}\otimes \mathbf{3}= \mathbf{1}\oplus \mathbf{1'}\oplus \mathbf{1''}\oplus \mathbf{3}_S\oplus \mathbf{3}_A$, we can obtain five independent weight 4 modular forms,
\begin{eqnarray}
\nonumber Y^{(4)}_{\mathbf{3}}&=&(Y^{(2)}_{\mathbf{3}}Y^{(2)}_{\mathbf{3}})_{\mathbf{3}}=(Y_1^2-Y_2 Y_3,Y_3^2-Y_1 Y_2,Y_2^2-Y_1 Y_3)^{T}\,, \\
\nonumber Y^{(4)}_{\mathbf{1}}&=&(Y^{(2)}_{\mathbf{3}}Y^{(2)}_{\mathbf{3}})_{\mathbf{1}}=Y_1^2+2 Y_2 Y_3\,, \\
Y^{(4)}_{\mathbf{1}'}&=&(Y^{(2)}_{\mathbf{3}}Y^{(2)}_{\mathbf{3}})_{\mathbf{1}'}=Y_3^2+2 Y_1 Y_2\,.
\end{eqnarray}
Similarly there are seven modular forms of weight 6, and they decompose into a singlet $\mathbf{1}$ and two triplets $\mathbf{3}$ under $A_4$,
\begin{eqnarray}
Y^{(6)}_{\mathbf{1}}&=&(Y^{(2)}_{\mathbf{3}}Y^{(4)}_{\mathbf{3}})_{\mathbf{1}}=Y_1^3+Y_2^3+Y_3^3-3 Y_1 Y_2 Y_3\,,\nonumber\\
Y^{(6)}_{\mathbf{3},1}&=&Y^{(2)}_{\mathbf{3}}Y^{(4)}_{\mathbf{1}}=(Y_1^3+2 Y_1Y_2Y_3,Y_1^2Y_2+2Y_2^2Y_3,Y_1^2Y_3+2Y_3^2Y_2)^{T}\,,\nonumber\\
Y^{(6)}_{\mathbf{3},2}&=&Y^{(2)}_{\mathbf{3}}Y^{(4)}_{\mathbf{1}'}=(Y_3^3+2 Y_1Y_2Y_3,Y_3^2Y_1+2Y_1^2Y_2,Y_3^2Y_2+2Y_2^2Y_1)^{T}\,.
\end{eqnarray}
Notice that $(Y^{(2)}_{\mathbf{3}}Y^{(2)}_{\mathbf{3}})_{\mathbf{1}''}$ is vanishing as shown in~Eq.\eqref{eq:MF2}.

\section{\label{sec:A4models}Neutrino mass models based on $\Gamma_3$ modular symmetry }

In this section,  we shall perform a systematical classification of all minimal neutrino mass models with the $\Gamma_3$ modular symmetry.
We adopt the $N=1$ global supersymmetry, the most general form of the action can be written as~\cite{Feruglio:2017spp}
\begin{equation}
\mathcal{S}=\int d^4 x d^2\theta d^2\bar \theta K(\Phi_I,\bar{\Phi}_I; \tau,\bar{\tau})+\int d^4 x d^2\theta W(\Phi_I,\tau)+\mathrm{h.c.}\,,
\end{equation}
where $K(\Phi_I,\bar{\Phi}_I, \tau,\bar{\tau})$ is the K$\ddot{\mathrm{a}}$hler potential, and $W$ denotes the superpotential. $\Phi_I$ is set of chiral superfields, under the modular transformation of Eq.~\eqref{eq:modular_trans}, it transforms as
\begin{equation}
\label{eq:modularTrs_Phi}
\tau\to \gamma\tau=\frac{a\tau+b}{c\tau+d}\,,\qquad
\Phi_I\to (c\tau+d)^{-k_I}\rho_I(\gamma)\Phi_I\,.
\end{equation}
where $-k_I$ is the modular weight, and $\rho_I(\gamma)$ is the unitary representation of the representative element $\gamma$ in $\Gamma_N$. There are no restrictions on the possible value of $k_I$ since the supermultiplets $\Phi_I$ are not modular forms. The K$\ddot{\mathrm{a}}$hler potential should be invariant
up to K$\ddot{\mathrm{a}}$hler transformations under the modular transformation of Eq.~\eqref{eq:modularTrs_Phi}. We shall use the following K$\ddot{\mathrm{a}}$hler potential in this work~\cite{Feruglio:2017spp},
\begin{equation}
K(\Phi_I,\bar{\Phi}_I; \tau,\bar{\tau}) =-h \log(-i\tau+i\bar\tau)+ \sum_I (-i\tau+i\bar\tau)^{-k_I} |\Phi_I|^2~~~,
\end{equation}
where $h$ is a positive constant $h>0$. After the modulus $\tau$ gets a vacuum expectation value (VEV), the above K$\ddot{\mathrm{a}}$hler potential leads to the following kinetic term for the scalar components of the supermultiplets $\Phi_I$ and the modulus superfield $\tau$,
\begin{equation}
\frac{h}{\langle-i\tau+i\bar\tau\rangle^2}\partial_\mu \bar\tau\partial^\mu \tau+\sum_I \frac{\partial_\mu  \bar{\phi}_I \partial^\mu \phi_I}{\langle-i\tau+i\bar\tau\rangle^{k_I}}\,.
\end{equation}
For a given value of the VEV of $\tau$, the kinetic term of $\phi_I$ can be made into canonical form by rescaling the fields $\phi_I$. This effect can be absorbed into the unknown free parameters of the superpotential in a specific model.

The superpotential $W(\Phi_I,\tau)$ can be expanded in power series of the involved supermultiplets $\Phi_I$,
\begin{equation}
W(\Phi_I,\tau) =\sum_n Y_{I_1...I_n}(\tau)~ \Phi_{I_1}... \Phi_{I_n}\,,
\end{equation}
where $Y_{I_1...I_n}$ is a modular multiplet of weight $k_Y$ and it transforms as the presentation $\rho_{Y}$ of $\Gamma_{N}$,
\begin{equation}
\tau\to \gamma\tau =\dfrac{a\tau+b}{c\tau+d}\,,\qquad Y(\tau)\to Y(\gamma\tau)=(c\tau+d)^{k_Y}\rho_{Y}(\gamma)Y(\tau)\,.
\end{equation}
The requirement of modular invariance of the superpotential implies
\begin{equation}
k_Y=k_{I_1}+...+k_{I_n},~\quad~ \rho_Y\otimes\rho_{I_1}\otimes\ldots\otimes\rho_{I_n}\ni\mathbf{1}\,.
\end{equation}
Then we proceed to discuss all possible simplest models for lepton masses and mixing with the $A_4$ modular symmetry. In order to construct models with the smallest number of free parameters, we don't introduce any flavon field other than the modulus $\tau$. The Higgs doublets $H_u$ and $H_d$ are assumed to transform as $\mathbf{1}$ under $A_4$ and their modular weights $k_{H_u, H_d}$ are vanishing. We consider two scenarios where the neutrino masses arise from the Weinberg operator and the type I seesaw mechanism. Similar to previous analyses~\cite{Feruglio:2017spp}, we assign the three generations of left-handed lepton doublets $L\equiv(L_1, L_2, L_3)^{T}$ and of the right-handed neutrino $N^c\equiv(N^c_1, N^c_2, N^c_3)^{T}$ to two triplets $\mathbf{3}$ of $A_4$ with modular weights denoted as $k_L$ and $k_{N^c}$. Unlike previous work~\cite{Feruglio:2017spp,Criado:2018thu,Kobayashi:2018scp,Okada:2018yrn,Kobayashi:2018wkl,Kobayashi:2018vbk,Novichkov:2018yse,Nomura:2019yft}, we allow right-handed charged leptons $E^{c}_{1, 2, 3}$ to transform as all combinations of $\mathbf{1}$, $\mathbf{1}'$ and $\mathbf{1}''$ representations of $A_4$, using the simplest different modular weights $k_{E_{1, 2, 3}}$ to break the degeneracy, leading to ten different charged lepton Yukawa matrices, instead of the usual one.

\subsection{\label{subsec:charged_lepton}Charged lepton sector}

Firstly we investigate the charged lepton sector. Since we do not allow any flavons (beyond the single modulus field $\tau$), we shall not attempt to explain the charged lepton mass hierarchy, which remains a challenge for modular symmetry models.
In order to avoid a charged lepton mass matrix with rank less than 3,  when two or all of $E^c_1,\,E^c_2$ and $E^c_3$ have same representation of $\Gamma_3$ , we assume that $E^c_1,\,E^c_2$ and $E^c_3$ have different modular weights such that they are distinguishable. For simplicity, we use lower weight modular forms as much as possible. Hence the model in the charged lepton sector can be divided into three possible cases.
\begin{itemize}[labelindent=-1.8em, leftmargin=1.8em]
\item[(i)]{$\rho_{E^{c}_1}=\rho_{E^{c}_2}=\rho_{E^{c}_3}$}

When all the three right-handed charged leptons $E^{c}_{1,2,3}$ transform as the same irreducible representation of $\Gamma_3$, they should carry different modular weights to distinguish from each other. As a consequence, the charged leptons $E^{c}_{1,2,3}$ could couple with the modular forms $Y^{(2)}_{\mathbf{3}}$, $Y^{(4)}_{\mathbf{3}}$ and $Y^{(6)}_{\mathbf{3}}$ respectively, and the superpotential for the charged lepton masses can be written as:
\begin{equation}
\label{eq:WeI}
W_e\,= \alpha(E^c_1 L Y^{(2)}_{\mathbf{3}})_\mathbf{1}H_d +\beta (E^c_2L Y^{(4)}_{\mathbf{3}})_\mathbf{1}H_d+\gamma(E^c_3LY^{(6)}_{\mathbf{3}})_\mathbf{1}H_d\,.
\end{equation}
The condition of modular invariance requires
\begin{equation}
k_{E_1}=k_{E_2}-2=k_{E_3}-4=2-k_L\,.
\end{equation}

\item[(ii)]{$\rho_{E^{c}_1}=\rho_{E^{c}_2}\neq\rho_{E^{c}_3}$}

If two of the three right-handed charged leptons $E^{c}_i$ transform in the same way under $A_4$\footnote{It is irrelevant that which two of the right-handed charged leptons share the same $A_4$ representation. Because this amounts to a row permutation of the charged lepton matrix $M_e$ in the right-left basis $E^cM_eL$, and the results for lepton mixing matrix is not changed. We shall choose $\rho_{E^{c}_1}=\rho_{E^{c}_2}$ for this case hereinafter.}, they could be assigned to different modular weights which are compensated by the lower weight modular forms $Y^{(2)}_{\mathbf{3}}$ and $Y^{(4)}_{\mathbf{3}}$. Thus the superpotential for the charged lepton masses are given by,
\begin{equation}
\label{eq:WeII}
W_e=\alpha(E^c_1LY^{(2)}_{\mathbf{3}})_\mathbf{1}H_d+\beta (E^c_2L Y^{(4)}_{\mathbf{3}})_\mathbf{1}H_d+ \gamma(E^c_3LY^{(2)}_{\mathbf{3}})_\mathbf{1}H_d\,,
\end{equation}
where the condition of weight cancellation entails
\begin{equation}
k_{E_1}=k_{E_2}-2=k_{E_3}=2-k_L\,.
\end{equation}

\item[(iii)]{$\rho_{E^{c}_1}\neq\rho_{E^{c}_2}\neq\rho_{E^{c}_3}$}

When the three right-handed charged leptons $E^{c}_i$ are assigned to
three different singlets $\mathbf{1}$, $\mathbf{1}'$ and $\mathbf{1}''$ of $A_4$ as in previous works~\cite{Feruglio:2017spp,Criado:2018thu,Kobayashi:2018scp,Okada:2018yrn,Kobayashi:2018wkl,Kobayashi:2018vbk,Novichkov:2018yse,Nomura:2019yft}, their modular weights could be identical, and only the lowest weight modular form $Y^{(2)}_{\mathbf{3}}$ is necessary in the minimal model. Then the superpotential for the charged lepton masses takes the form
\begin{equation}
\label{eq:WeIII}
W_e=\alpha(E^c_1LY^{(2)}_{\mathbf{3}})_\mathbf{1}H_d +\beta (E^c_2 L Y^{(2)}_{\mathbf{3}})_\mathbf{1}H_d + \gamma (E^c_3LY^{(2)}_{\mathbf{3}})_\mathbf{1}H_d\,.
\end{equation}
The invariance of $W_e$ under modular transformations implies the
following relations for the weights,
\begin{equation}
k_{E_1}=k_{E_2}=k_{E_3}=2-k_L\,.
\end{equation}

\end{itemize}

To be more specific, making use of the Clebsch-Gordan coefficients given in Eq.~\eqref{eq:CG_coefficient}, we can expand the superpotentials of  Eqs.~(\ref{eq:WeI}, \ref{eq:WeII}, \ref{eq:WeIII}) into the following forms for all possible singlet assignments of right-handed charged leptons. \begin{itemize}[labelindent=-1.8em, leftmargin=1.8em]
\item {$\rho_{E^c_{1,2,3}}=\mathbf{1},\,k_{E^c_{1,2,3}}+k_L=2,4,6$}
\begin{align}
\nonumber
W_e
&=\alpha E^c_1(LY^{(2)}_{\mathbf{3}})_\mathbf{1}H_d
+\beta E^c_2(LY^{(4)}_{\mathbf{3}})_\mathbf{1}H_d + \gamma_1 E^c_3(LY^{(6)}_{\mathbf{3},1})_\mathbf{1}H_d + \gamma_2 E^c_3(LY^{(6)}_{\mathbf{3},2})_\mathbf{1}H_d\\
\nonumber
&=\alpha E^c_1(L_1 Y_1+L_2 Y_3+L_3 Y_2)H_d\\
\nonumber
&\quad+\beta E^c_2[L_1(Y^2_1-Y_2Y_3)\,+\,L_2(Y^2_2-Y_1Y_3)+L_3(Y^2_3-Y_1Y_2)]H_d\\
\nonumber
&\quad+\gamma_1 E^c_3[L_1(Y^3_1+2Y_1Y_2Y_3)\,+\,L_2(Y^2_1Y_3+2Y^2_3Y_2)\,+\,L_3(Y^2_1Y_2+2Y^2_2Y_3)]H_d\\
\label{eq:We_1}
&\quad+\gamma_2 E^c_3[L_1(Y^3_3+2Y_1Y_2Y_3)\,+\,L_2(Y^2_3Y_2+2Y^2_2Y_1)\,+\,L_3(Y^2_3Y_1+2Y^2_1Y_2)]H_d\,.
\end{align}

\item {$\rho_{E^c_{1, 2, 3}}=\mathbf{1}',  k_{E^c_{1,2,3}}+k_L=2,4,6$}
\begin{align}
\nonumber
W_e
&=\alpha E^c_1(LY^{(2)}_{\mathbf{3}})_{\mathbf{1}''}H_d
+\beta E^c_2(LY^{(4)}_{\mathbf{3}})_{\mathbf{1}''}H_d + \gamma_1 E^c_3(LY^{(6)}_{\mathbf{3},1})_{\mathbf{1}''}H_d + \gamma_2 E^c_3(LY^{(6)}_{\mathbf{3},2})_{\mathbf{1}''}H_d\\
\nonumber
&=\alpha E^c_1(L_2 Y_2+L_3 Y_1+L_1 Y_3)H_d\\
\nonumber
&\quad+\beta E^c_2[L_2(Y^2_3-Y_1Y_2)\,+\,L_3(Y^2_1-Y_2Y_3)+L_1(Y^2_2-Y_1Y_3)]H_d\\
\nonumber
&\quad+\gamma_1 E^c_3[L_2(Y^2_1Y_2+2Y^2_2Y_3)\,+\,L_3(Y^3_1+2Y_1Y_2Y_3)\,+\,L_1(Y^2_1Y_3+2Y^2_3Y_2)]H_d\\
\label{eq:We_2}
&\quad+\gamma_2 E^c_3[L_2(Y^2_3Y_1+2Y^2_1Y_2)\,+\,L_3(Y^3_3+2Y_1Y_2Y_3)\,+\,L_1(Y^2_3Y_2+2Y^2_2Y_1)]H_d\,.
\end{align}

\item {$\rho_{E^c_{1,2,3}} =\mathbf{1}'',\,k_{E^c_{1,2,3}}+k_L=2,4,6$}
\begin{align}
\nonumber
W_e
&=\alpha E^c_1(LY^{(2)}_{\mathbf{3}})_\mathbf{1'}H_d
+\beta E^c_2(LY^{(4)}_{\mathbf{3}})_\mathbf{1'}H_d + \gamma_1 E^c_3(LY^{(6)}_{\mathbf{3},1})_\mathbf{1'}H_d + \gamma_2 E^c_3(LY^{(6)}_{\mathbf{3},2})_\mathbf{1'}H_d\\
\nonumber
&=\alpha E^c_1(L_3\,Y_3+L_1\,Y_2+L_2\,Y_1)H_d\\
\nonumber
&\quad+\beta E^c_2[L_3(Y^2_2-Y_1Y_3)+L_1(Y^2_3-Y_1Y_2)+L_2(Y^2_1-Y_2Y_3)]H_d\\
\nonumber
&\quad+\gamma_1 E^c_3[L_3(Y^2_1Y_3+2Y^2_3Y_2)+L_1(Y^2_1Y_2+2Y^2_2Y_3)+L_2(Y^3_1+2Y_1Y_2Y_3)]H_d\\
&\quad+\gamma_2 E^c_3[L_3(Y^2_3Y_2+2Y^2_2Y_1)+L_1(Y^2_3Y_1+2Y^2_1Y_2)+L_2(Y^3_3+2Y_1Y_2Y_3)]H_d\,.
\label{eq:We_3}
\end{align}

\item {$\rho_{E^c_{1,2,3}}=\mathbf{1}, \mathbf{1}, \mathbf{1}'$,~$k_{E^c_{1,2,3}}+k_L=2,4,2$}
\begin{align}
\nonumber
W_e
&=\alpha E^c_1(LY^{(2)}_{\mathbf{3}})_\mathbf{1}H_d
+\beta E^c_2(L\,Y^{(4)}_{\mathbf{3}})_\mathbf{1}H_d + \gamma E^c_3(LY^{(2)}_{\mathbf{3}})_{\mathbf{1}''}H_d\\
\nonumber
&=\alpha E^c_1(L_1 Y_1+L_2 Y_3+L_3 Y_2)H_d+\beta E^c_2\big[L_1(Y^2_1-Y_2Y_3)+L_2(Y^2_2-Y_1Y_3)\\
&\quad+L_3(Y^2_3-Y_1Y_2)\big]H_d+\gamma E^c_3(L_2Y_2+L_3Y_1+L_1 Y_3)H_d\,.
\label{eq:We_4}
\end{align}

\item {$\rho_{E^c_{1,2,3}}=\mathbf{1}, \mathbf{1}, \mathbf{1}''$,~ $k_{E^c_{1,2,3}}+k_L=2, 4, 2$}
\begin{align}
\nonumber
W_e
&=\alpha E^c_1(LY^{(2)}_{\mathbf{3}})_\mathbf{1}H_d
+\beta E^c_2(L\,Y^{(4)}_{\mathbf{3}})_\mathbf{1}H_d + \gamma E^c_3(LY^{(2)}_{\mathbf{3}})_\mathbf{1'}H_d\\
\nonumber
&=\alpha E^c_1(L_1 Y_1+L_2 Y_3+L_3 Y_2)H_d+\beta E^c_2\big[L_1(Y^2_1-Y_2Y_3)+L_2(Y^2_2-Y_1Y_3)\\
&\quad+L_3(Y^2_3-Y_1Y_2)\big]H_d+\gamma E^c_3(L_3 Y_3+L_1 Y_2+L_2 Y_1)H_d\,.
\label{eq:We_5}
\end{align}

\item {$\rho_{E^c_{1,2,3}}=\mathbf{1}', \mathbf{1}', \mathbf{1}$,~ $k_{E^c_{1,2,3}}+k_L=2,4,2$}
\begin{align}
\nonumber
W_e
&=\alpha E^c_1(LY^{(2)}_{\mathbf{3}})_{\mathbf{1}''}H_d
+\beta E^c_2(LY^{(4)}_{\mathbf{3}})_{\mathbf{1}''}H_d + \gamma E^c_3(LY^{(2)}_{\mathbf{3}})_\mathbf{1}H_d\\
\nonumber
&=\alpha E^c_1(L_2 Y_2+L_3 Y_1+L_1 Y_3)H_d+\beta E^c_2\big[L_2(Y^2_3-Y_1Y_2)+L_3(Y^2_1-Y_2Y_3)\\
&\quad+L_1(Y^2_2-Y_1Y_3)\big]H_d +\gamma E^c_3(L_1 Y_1+L_2 Y_3+L_3 Y_2)H_d\,.
\label{eq:We_6}
\end{align}

\item {$\rho_{E^c_{1,2,3}}=\mathbf{1}', \mathbf{1}', \mathbf{1}''$,~ $k_{E^c_{1,2,3}}+k_L=2,4,2$}
\begin{align}
\nonumber
W_e
&=\alpha E^c_1(LY^{(2)}_{\mathbf{3}})_{\mathbf{1}''}H_d
+\beta E^c_2(L\,Y^{(4)}_{\mathbf{3}})_{\mathbf{1}''}H_d + \gamma E^c_3(LY^{(2)}_{\mathbf{3}})_{\mathbf{1}'}H_d\\
\nonumber
&=\alpha E^c_1(L_2 Y_2+L_3 Y_1+L_1 Y_3)H_d+\beta E^c_2\big[L_2(Y^2_3-Y_1Y_2)+L_3(Y^2_1-Y_2Y_3)\\
&\quad+L_1(Y^2_2-Y_1Y_3)\big]H_d+\gamma E^c_3(L_3 Y_3+L_1 Y_2\,+\,L_2 Y_1)H_d\,.
\label{eq:We_7}
\end{align}

\item {$\rho_{E^c_{1,2,3}}=\mathbf{1}'', \mathbf{1}'', \mathbf{1}$,~ $k_{E^c_{1,2,3}}+k_L=2, 4, 2$ }
\begin{align}
\nonumber
W_e
&=\alpha E^c_1(LY^{(2)}_{\mathbf{3}})_{\mathbf{1}'}H_d
+\beta E^c_2(LY^{(4)}_{\mathbf{3}})_{\mathbf{1}'}H_d + \gamma E^c_3(LY^{(2)}_{\mathbf{3}})_\mathbf{1}H_d\\
\nonumber
&=\alpha E^c_1(L_3 Y_3+L_1 Y_2+L_2 Y_1)H_d+\beta E^c_2\big[L_3(Y^2_2-Y_1Y_3)+L_1(Y^2_3-Y_1Y_2)\\
&\quad+L_2(Y^2_1-Y_2Y_3)\big]H_d+\gamma E^c_3(L_1 Y_1+L_2 Y_3+L_3 Y_2)H_d\,.
\label{eq:We_8}
\end{align}

\item {$\rho_{E^c_{1,2,3}}=\mathbf{1}'', \mathbf{1}'', \mathbf{1}'$,~ $k_{E^c_{1,2,3}}+k_L=2, 4, 2$}
\begin{align}
\nonumber
W_e
&=\alpha E^c_1(LY^{(2)}_{\mathbf{3}})_{\mathbf{1}'}H_d
+\beta E^c_2(LY^{(4)}_{\mathbf{3}})_{\mathbf{1}'}H_d + \gamma E^c_3(LY^{(2)}_{\mathbf{3}})_{\mathbf{1}''}H_d\\
\nonumber
&=\alpha E^c_1(L_3 Y_3+L_1 Y_2+L_2 Y_1)H_d+\beta E^c_2\big[L_3(Y^2_2-Y_1Y_3)+L_1(Y^2_3-Y_1Y_2)\\
&\quad+L_2(Y^2_1-Y_2Y_3)\big]H_d+\gamma E^c_3(L_2 Y_2+L_3 Y_1+L_1 Y_3)H_d\,.
\label{eq:We_9}
\end{align}

\item {$\rho_{E^c_{1,2,3}}=\mathbf{1}, \mathbf{1}'', \mathbf{1}'$,~ $k_{E^c_{1,2,3}}+k_L=2, 2, 2$}
\begin{align}
\nonumber
W_e
&=\alpha E^c_1(LY^{(2)}_{\mathbf{3}})_{\mathbf{1}}H_d
+\beta E^c_2(LY^{(2)}_{\mathbf{3}})_{\mathbf{1}'}H_d + \gamma E^c_3(LY^{(2)}_{\mathbf{3}})_{\mathbf{1}''}H_d\\
\nonumber
&=\alpha E^c_1(L_1 Y_1+L_2 Y_3+L_3 Y_2)H_d+\beta E^c_2(L_3 Y_3+L_1 Y_2+L_2 Y_1)H_d\\
&\quad+\gamma E^c_3(L_2 Y_2+L_3 Y_1+L_1 Y_3)H_d\,.
\label{eq:We_10}
\end{align}
This is exactly the original $A_4$ modular symmetry model considered in the literature~\cite{Feruglio:2017spp,Criado:2018thu,Kobayashi:2018scp,Okada:2018yrn,Kobayashi:2018wkl,Kobayashi:2018vbk,Novichkov:2018yse,Nomura:2019yft}. 
The resulting charged lepton mass matrices for each possible model considered above are summarized in table~\ref{tab:charged lepton}.

\end{itemize}

\begin{table}[t!]
\renewcommand{\tabcolsep}{0.58mm}
\centering
\begin{tabular}{|c|c|c|c|} \hline\hline
 & $\rho_{E^c_{1,2,3}}$ & $k_{E^c_{1,2,3}}+k_L$ & Charged lepton mass matrices	\\ \hline
 & & & \\[-0.15in]
$C_1$  & $\mathbf{1}$, $\mathbf{1}$, $\mathbf{1}$ & 2, 4, 6 & {\scriptsize $ M_e = \begin{pmatrix}
	 ~~&~~     ~~&~~    ~~&~~ \\[-0.1in]
 \alpha\,Y_1 ~~&~~ \alpha\,Y_3 ~~&~~ \alpha\,Y_2 \\
 ~~&~~     ~~&~~    ~~&~~ \\[-0.02in]
\beta(Y_1^2-Y_2Y_3) ~&~ \beta(Y_2^2-Y_1Y_3)  ~&~ \beta(Y_3^2-Y_1Y_2) \\
 ~&~        ~&~      ~&~  \\[-0.02in]
\gamma_1(Y_1^3+2Y_1Y_2Y_3) ~&~ \gamma_1(Y_1^2Y_3+2Y_3^2Y_2) ~&~\gamma_1(Y_1^2Y_2+2Y_2^2Y_3)\\
+\gamma_2(Y_3^3+2Y_1Y_2Y_3) ~&~    +\gamma_2(Y_3^2Y_2+2Y_2^2Y_1)  ~&~
+\gamma_2(Y_3^2Y_1+2Y_1^2Y_2)\\
 ~~&~~     ~~&~~    ~~&~~ \\[-0.1in]
 \end{pmatrix}v_d $}\\
 & & & \\[-0.15in]  \hline
  & & & \\[-0.15in]
$C_2$  & $\mathbf{1}'$, $\mathbf{1}'$, $\mathbf{1}'$  & 2, 4, 6  &  {\scriptsize $ M_e = \begin{pmatrix}
	 ~~&~~     ~~&~~    ~~&~~ \\[-0.1in]
 \alpha\,Y_3 ~~&~~ \alpha\,Y_2 ~~&~~ \alpha\,Y_1 \\
 ~~&~~     ~~&~~    ~~&~~ \\[-0.02in]
\beta(Y_2^2-Y_1Y_3) ~&~ \beta(Y_3^2-Y_1Y_2)  ~&~ \beta(Y_1^2-Y_2Y_3) \\
 ~&~        ~&~      ~&~  \\[-0.02in]
 \gamma_1(Y_1^2Y_3+2Y_3^2Y_2)~&~ \gamma_1(Y_1^2Y_2+2Y_2^2Y_3) ~&~\gamma_1(Y_1^3+2Y_1Y_2Y_3)\\
+\gamma_2(Y_3^2Y_2+2Y_2^2Y_1) ~&~ +\gamma_2(Y_3^2Y_1+2Y_1^2Y_2)  ~&~
+\gamma_2(Y_3^3+2Y_1Y_2Y_3)\\
 ~~&~~     ~~&~~    ~~&~~ \\[-0.1in]
 \end{pmatrix} v_d$}\\
 & & & \\[-0.15in]  \hline
  & & & \\[-0.15in]
$C_3$  & $\mathbf{1}''$, $\mathbf{1}''$, $\mathbf{1}''$  & 2, 4, 6  &  {\scriptsize $ M_e =\begin{pmatrix}
	 ~~&~~     ~~&~~    ~~&~~ \\[-0.1in]
 \alpha\,Y_2 ~~&~~ \alpha\,Y_1 ~~&~~ \alpha\,Y_3 \\
 ~~&~~     ~~&~~    ~~&~~ \\[-0.02in]
\beta(Y_3^2-Y_1Y_2) ~&~ \beta(Y_1^2-Y_2Y_3)  ~&~ \beta(Y_2^2-Y_1Y_3) \\
 ~&~        ~&~      ~&~  \\[-0.02in]
 \gamma_1(Y_1^2Y_2+2Y_2^2Y_3)~&~ \gamma_1(Y_1^3+2Y_1Y_2Y_3) ~&~\gamma_1(Y_1^2Y_3+2Y_3^2Y_2)\\
+\gamma_2(Y_3^2Y_1+2Y_1^2Y_2) ~&~ +\gamma_2(Y_3^3+2Y_1Y_2Y_3) ~&~ +\gamma_2(Y_3^2Y_2+2Y_2^2Y_1)\\
 ~~&~~     ~~&~~    ~~&~~ \\[-0.1in]
 \end{pmatrix}v_d $}\\
  & & & \\[-0.15in]  \hline
  & & & \\[-0.15in]
$C_4$  & $\mathbf{1}$, $\mathbf{1}$, $\mathbf{1}'$  & 2, 4, 2  &  {\scriptsize $ M_e = \begin{pmatrix}
	 ~~&~~     ~~&~~    ~~&~~ \\[-0.1in]
 \alpha\,Y_1 ~~&~~ \alpha\,Y_3 ~~&~~ \alpha\,Y_2 \\
 ~~&~~     ~~&~~    ~~&~~ \\[-0.02in]
\beta(Y_1^2-Y_2Y_3) ~&~ \beta(Y_2^2-Y_1Y_3)  ~&~ \beta(Y_3^2-Y_1Y_2) \\
 ~&~        ~&~      ~&~  \\[-0.02in]
 \gamma Y_3~&~ \gamma Y_2~&~\gamma Y_1\\
 ~~&~~     ~~&~~    ~~&~~ \\[-0.1in]
 \end{pmatrix} v_d $}\\
  & & & \\[-0.15in]  \hline
  & & & \\[-0.15in]
 $C_5$  & $\mathbf{1}$, $\mathbf{1}$, $\mathbf{1}''$  & 2, 4, 2  &  {\scriptsize $ M_e =\begin{pmatrix}
	 ~~&~~     ~~&~~    ~~&~~ \\[-0.1in]
 \alpha\,Y_1 ~~&~~ \alpha\,Y_3 ~~&~~ \alpha\,Y_2 \\
 ~~&~~     ~~&~~    ~~&~~ \\[-0.02in]
\beta(Y_1^2-Y_2Y_3) ~&~ \beta(Y_2^2-Y_1Y_3)  ~&~ \beta(Y_3^2-Y_1Y_2) \\
 ~&~        ~&~      ~&~  \\[-0.02in]
 \gamma Y_2~&~ \gamma Y_1~&~\gamma Y_3\\
 ~~&~~     ~~&~~    ~~&~~ \\[-0.1in]
 \end{pmatrix}v_d $}\\
   & & & \\[-0.15in]  \hline
   & & & \\[-0.15in]	
$C_6$  & $\mathbf{1}'$, $\mathbf{1}'$, $\mathbf{1}$  & 2, 4, 2  &  {\scriptsize $ M_e =\begin{pmatrix}
	 ~~&~~     ~~&~~    ~~&~~ \\[-0.1in]
 \alpha\,Y_3 ~~&~~ \alpha\,Y_2 ~~&~~ \alpha\,Y_1 \\
 ~~&~~     ~~&~~    ~~&~~ \\[-0.02in]
\beta(Y_2^2-Y_1Y_3) ~&~ \beta(Y_3^2-Y_1Y_2)  ~&~ \beta(Y_1^2-Y_2Y_3) \\
 ~&~        ~&~      ~&~  \\[-0.02in]
 \gamma Y_1~&~ \gamma Y_3~&~\gamma Y_2\\
 ~~&~~     ~~&~~    ~~&~~ \\[-0.1in]
 \end{pmatrix} v_d $}\\
  & & & \\[-0.15in]  \hline
  & & & \\[-0.15in]	
 $C_7$  & $\mathbf{1}'$, $\mathbf{1}'$, $\mathbf{1}''$  & 2, 4, 2   &  {\scriptsize $ M_e =\begin{pmatrix}
	 ~~&~~     ~~&~~    ~~&~~ \\[-0.1in]
 \alpha\,Y_3 ~~&~~ \alpha\,Y_2 ~~&~~ \alpha\,Y_1 \\
 ~~&~~     ~~&~~    ~~&~~ \\[-0.02in]
\beta(Y_2^2-Y_1Y_3) ~&~ \beta(Y_3^2-Y_1Y_2)  ~&~ \beta(Y_1^2-Y_2Y_3) \\
 ~&~        ~&~      ~&~  \\[-0.02in]
 \gamma Y_2~&~ \gamma Y_1~&~\gamma Y_3\\
 ~~&~~     ~~&~~    ~~&~~ \\[-0.1in]
 \end{pmatrix}v_d $}\\
  & & & \\[-0.15in]  \hline
  & & & \\[-0.15in]
 $C_8$  & $\mathbf{1}''$, $\mathbf{1}''$, $\mathbf{1}$  & 2, 4, 2  &  {\scriptsize $ M_e =\begin{pmatrix}
	 ~~&~~     ~~&~~    ~~&~~ \\[-0.1in]
 \alpha\,Y_2 ~~&~~ \alpha\,Y_1 ~~&~~ \alpha\,Y_3 \\
 ~&~     ~&~    ~&~ \\[-0.02in]
\beta(Y_3^2-Y_1Y_2) ~&~ \beta(Y_1^2-Y_2Y_3)  ~&~ \beta(Y_2^2-Y_1Y_3) \\
 ~&~        ~&~      ~&~  \\[-0.02in]
 \gamma Y_1~&~ \gamma Y_3~&~\gamma Y_2\\
 ~~&~~     ~~&~~    ~~&~~ \\[-0.1in]
 \end{pmatrix}v_d $}\\
  & & & \\[-0.15in]  \hline
  & & & \\[-0.15in]
 $C_9$  & $\mathbf{1}''$, $\mathbf{1}''$, $\mathbf{1}'$  & 2, 4, 2  &  {\scriptsize $ M_e =\begin{pmatrix}
	 ~~&~~     ~~&~~    ~~&~~ \\[-0.1in]
 \alpha\,Y_2 ~~&~~ \alpha\,Y_1 ~~&~~ \alpha\,Y_3 \\
 ~~&~~     ~~&~~    ~~&~~ \\[-0.02in]
\beta(Y_3^2-Y_1Y_2) ~&~ \beta(Y_1^2-Y_2Y_3)  ~&~ \beta(Y_2^2-Y_1Y_3) \\
 ~&~        ~&~      ~&~  \\[-0.02in]
 \gamma Y_3~&~ \gamma Y_2~&~\gamma Y_1\\
 ~~&~~     ~~&~~    ~~&~~ \\[-0.1in]
 \end{pmatrix} v_d $}\\
  & & & \\[-0.15in]  \hline
  & & & \\[-0.15in]	
 $C_{10}$  & $\mathbf{1}$, $\mathbf{1}''$, $\mathbf{1}'$  & 2, 2, 2  &  {\scriptsize $ M_e =\begin{pmatrix}
	 ~~&~~     ~~&~~    ~~&~~ \\[-0.1in]
 \alpha\,Y_1 ~~&~~ \alpha\,Y_3 ~~&~~ \alpha\,Y_2 \\
 ~~&~~     ~~&~~    ~~&~~ \\[-0.02in]
\beta Y_2 ~&~ \beta Y_1  ~&~ \beta Y_3 \\
 ~&~        ~&~      ~&~  \\[-0.02in]
 \gamma Y_3~&~ \gamma Y_2~&~\gamma Y_1\\
 ~~&~~     ~~&~~    ~~&~~ \\[-0.1in]
 \end{pmatrix} v_d $}\\  \hline	\hline				 	 		 		
\end{tabular}
\caption{\label{tab:charged lepton}The charged lepton mass matrices for different possible assignments of the right-handed charged leptons, where the charged lepton mass matrix $M_e$ is given in the right-left basis $E^c\,M_e\,L$ with $ v_d =\langle H^0_d\rangle$.  }
\end{table}

\subsection{\label{subsec:neutrino}Neutrino sector}

We don't know the nature of neutrinos which can be either Dirac particles similar to electron or Majorana particles. In this section, we shall consider the case of Majorana neutrinos, Dirac neutrinos can be analyzed in a similar manner. Thus the neutrino masses can arise from the effective Weinberg operator or the seesaw mechanism. In order to construct minimal models, we consider the cases that the complex modulus $\tau$ is involved through the lowest nontrivial weight 2 modular form $Y^{(2)}_{\mathbf{3}}$ in the following. If neutrino masses are described by the Weinberg operator and the three lepton doublets are assigned to an $A_4$ triplet $\mathbf{3}$, the simplest superpotential for neutrino masses is
\begin{equation}
W_\nu =  \frac{1}{\Lambda}\big(H_u H_u L L Y\big)_\mathbf{1}\,
=2\big[(L^2_1-L_2L_3)Y_1+(L^2_2-L_1L_3)Y_2+(L^2_3-L_1L_2)Y_3 \big]\frac{H^2_u}{\Lambda}\,.
\label{eq:Wnu_w1}
\end{equation}
Obviously the modular weight of the lepton doublet $L$ should be $k_L=1$ in this case. The resulting prediction for the neutrino mass matrix is
\begin{equation}
 M_\nu = \begin{pmatrix}
 2Y_1 ~&~ -Y_3 ~&~ -Y_2 \\
-Y_3 ~&~ 2Y_2  ~&~ -Y_1  \\
-Y_2 ~&~ -Y_1 ~&~2Y_3
\end{pmatrix}\dfrac{v^2_u}{\Lambda}\,.
\end{equation}
If neutrino masses are generated through the type-I seesaw mechanism, for the triplet assignments of both right-handed neutrinos $N^{c}$ and left-handed lepton doublets $L$, the most general form of the superpotential in the neutrino sector is
\begin{equation}
W_\nu = g \left(N^c L H_uf_N\left(Y\right)\right)_\mathbf{1}
+ \Lambda \left(N^c N^cf_M\left(Y\right)\right)_\mathbf{1}\,,
\label{eq:WnuII}
\end{equation}
where $f_N(Y)$ and $f_M(Y)$ are generic functions of the modular forms $Y(\tau)$. Motivated by the principle of minimality, we consider the
cases that $f_N(Y)$ and $f_M(Y)$ are either constant or proportional to $Y^{(2)}_{\mathbf{3}}$. Then we have the following three possible cases.

\begin{itemize}[labelindent=-1.8em, leftmargin=1.8em]

\item{$f_N\left(Y\right)\propto Y^{(2)}_{\mathbf{3}}$ and $f_M\left(Y\right)\propto 1$ }
\begin{align}
\nonumber W_\nu
&=g_1((N^c\,L)_{\mathbf{3}_S}Y^{(2)}_{\mathbf{3}})_\mathbf{1}H_u+g_2((N^cL)_{\mathbf{3}_A}Y^{(2)}_{\mathbf{3}})_\mathbf{1}H_u\,
+\Lambda \left(N^c N^c\right)_\mathbf{1}\\
\nonumber
&=g_1\big[(2N^c_1\,L_1\,-N^c_2\,L_3-N^c_3\,L_2)Y_1+(2N^c_3\,L_3\,-N^c_1\,L_2-N^c_2\,L_1)Y_3 \\
\nonumber
&\quad +(2N^c_2L_2-N^c_1L_3-N^c_3L_1)Y_2\big]H_u+g_2\big[(N^c_2L_3-N^c_3L_2)Y_1\\
&\quad +(N^c_1L_2-N^c_2L_1)Y_3+(N^c_3L_1-N^c_1L_3)Y_2\big]H_u+\Lambda(N^c_1N^c_1+2N^c_2N^c_3)\,.
\label{eq:Wnu_s1}
\end{align}
In this case the weights of $N^c$ and $L$ should be $k_{N^c}=0, k_L=2$. The Dirac neutrino mass matrix and the right-handed neutrino heavy Majorana mass matrix read as
\begin{equation}
M_D = \begin{pmatrix}
2g_1Y_1        ~&~  (-g_1+g_2)Y_3 ~&~ (-g_1-g_2)Y_2 \\
(-g_1-g_2)Y_3  ~&~     2g_1Y_2    ~&~ (-g_1+g_2)Y_1  \\
 (-g_1+g_2)Y_2 ~&~ (-g_1-g_2)Y_1  ~&~ 2g_1Y_3
\end{pmatrix}v_{u},~~~~
M_N =\begin{pmatrix}
1 ~& 0 ~& 0 \\
0 ~& 0 ~& 1 \\
0 ~& 1 ~& 0
\end{pmatrix}\Lambda \,,
\end{equation}
with $v_{u} \equiv \langle H^0_{u}\rangle$.

\item{$f_N\left(Y\right)\propto 1$ and $f_M\left(Y\right)\propto Y^{(2)}_{\mathbf{3}}$}

\begin{align}
\nonumber
W_\nu
&=g((N^c\,L)_\mathbf{1}H_u+\Lambda (\left(N^c\, N^c\right)_{\mathbf{3}_S}Y^{(2)}_{\mathbf{3}})_\mathbf{1}\\
\nonumber
&=g(N^c_1\,L_1\,+N^c_2\,L_3+N^c_3\,L_2)H_u+2\Lambda\big[(N^c_1N^c_1-N^c_2N^c_3)Y_1 \\
&\quad +(N^c_3N^c_3-N^c_1N^c_2)Y_3+(N^c_2N^c_2-N^c_1N^c_3)Y_2\big]\,.
\label{eq:Wnu_s2}
\end{align}
The condition of weight cancellation requires $k_{N^c}=-k_L=1$. We can read out the expressions of $M_D$ and $M_R$ as follow,
\begin{equation}
M_D = g\begin{pmatrix}
1 ~& 0 ~& 0 \\
0 ~& 0 ~& 1 \\
0 ~& 1 ~&0
\end{pmatrix}v_{u},\qquad
M_N = \begin{pmatrix}
2Y_1 ~&~ -Y_3 ~&~ -Y_2 \\
-Y_3 ~&~ 2Y_2  ~&~ -Y_1  \\
-Y_2 ~&~ -Y_1 ~&~2Y_3
\end{pmatrix} \Lambda\,.
\end{equation}

\begin{table}[t!]
\centering
\resizebox{1.1\textwidth}{!}{
\begin{tabular}{|c|c|c|} \hline\hline

& $k_L$, $k_{N^c}$ &	Neutrino mass matrices  \\ \hline
   &   &     \\ [-0.15in]	
$W_1$ & 1, ---  & {\scriptsize $ M_\nu = \begin{pmatrix}
 2Y_1 ~&~ -Y_3 ~&~ -Y_2 \\
-Y_3 ~&~ 2Y_2  ~&~ -Y_1  \\
-Y_2 ~&~ -Y_1 ~&~2Y_3 \\
\end{pmatrix}\dfrac{v^2_u}{\Lambda}$}  \\
  &  &    \\ [-0.15in] \hline
   &  &     \\ [-0.15in]
$S_1$ &  2, 0 &  {\scriptsize $ M_D = \begin{pmatrix}
2g_1Y_1        ~&~  (-g_1+g_2)Y_3 ~&~ (-g_1-g_2)Y_2 \\
(-g_1-g_2)Y_3  ~&~     2g_1Y_2    ~&~ (-g_1+g_2)Y_1  \\
 (-g_1+g_2)Y_2 ~&~ (-g_1-g_2)Y_1  ~&~ 2g_1Y_3 \\
 \end{pmatrix}v_{u} $},~~{\scriptsize	$M_N = \begin{pmatrix}
1 & 0 & 0 \\
0 & 0 & 1 \\
0 & 1 & 0 \\
\end{pmatrix}\Lambda $} \\

   &  &     \\ [-0.15in]\hline
   &  &    \\ [-0.15in]

$S_2$ & $-1$, 1  &  {\scriptsize $ M_D = g\begin{pmatrix}

1 & 0 & 0 \\
0 & 0 & 1 \\
0 & 1 &0 \\
\end{pmatrix}v_{u}$},~~{\scriptsize	$M_N = \begin{pmatrix}
  2Y_1 ~&~ -Y_3 ~&~ -Y_2 \\
 -Y_3 ~&~ 2Y_2  ~&~ -Y_1  \\
 -Y_2 ~&~ -Y_1 ~&~2Y_3 \\
\end{pmatrix} \Lambda$}  \\
 &  &     \\ [-0.15in] \hline

&  &     \\ [-0.15in]

$S_3$ & 1, 1  &   {\scriptsize $ M_D = \begin{pmatrix}
2g_1Y_1        ~&~  (-g_1+g_2)Y_3 ~&~ (-g_1-g_2)Y_2 \\
(-g_1-g_2)Y_3  ~&~     2g_1Y_2    ~&~ (-g_1+g_2)Y_1  \\
(-g_1+g_2)Y_2 ~&~ (-g_1-g_2)Y_1  ~&~ 2g_1Y_3 \\
\end{pmatrix}v_{u} $},~~{\scriptsize$M_N = \begin{pmatrix}
  2Y_1 ~&~ -Y_3 ~&~ -Y_2 \\
 -Y_3 ~&~ 2Y_2  ~&~ -Y_1  \\
 -Y_2 ~&~ -Y_1 ~&~2Y_3 \\
 ~&~ ~&~\\[-0.1in]
\end{pmatrix}\Lambda$}   \\ \hline \hline			 		
\end{tabular} }
\caption{\label{tab:neutrino}The predictions for the neutrino mass matrices, where we assume that only the lowest weight 2 modular forms are involved, and the neutrino masses are generated through the Weinberg operator for $W_1$ and the type-I seesaw mechanism for the models $S_{1,2,3}$. }
\end{table}

\item{$f_N\left(Y\right)\propto Y^{(2)}_{\mathbf{3}}$ and $f_M\left(Y\right)\propto Y^{(2)}_{\mathbf{3}}$   }

\begin{align}
\nonumber
W_\nu
&=g_1((N^c\,L)_{\mathbf{3}_S}Y^{(2)}_{\mathbf{3}})_\mathbf{1}H_u+g_2((N^c\,L)_{\mathbf{3}_A}Y^{(2)}_{\mathbf{3}})_\mathbf{1}H_u
+\Lambda (\left(N^c N^c\right)_\mathbf{3_S}Y)_\mathbf{1}\\
\nonumber
&=g_1\big[(2N^c_1L_1-N^c_2L_3-N^c_3L_2)Y_1+(2N^c_3L_3-N^c_1L_2-N^c_2L_1)Y_3 \\
\nonumber
&~~+(2N^c_2L_2-N^c_1L_3-N^c_3L_1)Y_2\big]H_u+g_2\big[(N^c_2L_3-N^c_3L_2)Y_1+(N^c_1L_2-N^c_2L_1)Y_3\\
\nonumber&~~+(N^c_3L_1-N^c_1L_3)Y_2\big]H_u+2\Lambda\big[(N^c_1N^c_1-N^c_2N^c_3)Y_1+(N^c_3N^c_3-N^c_1N^c_2)Y_3 \\
&~~+(N^c_2N^c_2-N^c_1N^c_3)Y_2\big]\,.
\label{eq:Wnu_s3}
\end{align}
The modular weights of $N^c$ and $L$ should be $k_L=k_{N^c}=1$. We find $M_D$ and $M_N$ take the following form
\begin{eqnarray}
\nonumber&&\qquad\quad~~~ M_N = \begin{pmatrix}
2Y_1 ~&~ -Y_3 ~&~ -Y_2 \\
 -Y_3 ~&~ 2Y_2  ~&~ -Y_1  \\
 -Y_2 ~&~ -Y_1 ~&~2Y_3
\end{pmatrix}\Lambda\,,\\
&&
M_D =\begin{pmatrix}
2g_1Y_1        ~&~  (-g_1+g_2)Y_3 ~&~ (-g_1-g_2)Y_2 \\
(-g_1-g_2)Y_3  ~&~     2g_1Y_2    ~&~ (-g_1+g_2)Y_1  \\
 (-g_1+g_2)Y_2 ~&~ (-g_1-g_2)Y_1  ~&~ 2g_1Y_3
\end{pmatrix}v_{u}\,.
\end{eqnarray}

\end{itemize}
We listed the predicted neutrino mass matrices for the above four cases in table~\ref{tab:neutrino}. Taking into account the possible structures of the models in the charged lepton and neutrino sectors discussed in above, we find there are totaly forty minimal neutrino mass models based on the $A_4$ modular symmetry: ten different Weinberg models and thirty different type I seesaw models, these models are named as $\mathcal{A}_i$, $\mathcal{B}_i$, $\mathcal{C}_i$ and $\mathcal{D}_i$ $(i=1,\ldots,10)$. Notice that
the modular weights of the matter fields can be fixed uniquely in each model, and they are listed in table~\ref{tab:poss_models}.

\begin{table*}[t!]
\centering
\resizebox{1.05\textwidth}{!}{	
\begin{tabular}{|c|c|c||c|c|c|} \hline\hline
\multirow{2}{*}{Models} & \multirow{2}{*}{mass matrices} &\multirow{2}{*}{$A_4$} & \multicolumn{3}{|c|}{modular weights} \\ \cline{4-6}
 & & & $k_{E^c_{1,2,3}}$ &$k_L $ & $k_{N^c}$ \\ \hline
 $\mathcal{A}_1$ & $W_1, C_1$ & $\mathbf{1}$, $\mathbf{1}$, $\mathbf{1}$ & $1,3,5$ & $1$ & ---\\ \hline
 $\mathcal{A}_2$ & $W_1,C_2$ & $\mathbf{1}'$, $\mathbf{1}'$, $\mathbf{1}'$ & $1,3,5$ & $1$ & --- \\ \hline
 $\mathcal{A}_3$ & $W_1, C_3$ & $\mathbf{1}''$, $\mathbf{1}''$, $\mathbf{1}''$ & $1,3,5$ & $1$ & ---\\ \hline	
$\mathcal{A}_4$ & $W_1, C_4$ & $\mathbf{1}$, $\mathbf{1}$, $\mathbf{1}'$ & $1,3,1$ & $1$ & --- \\  \hline	
 $\mathcal{A}_5$ & $W_1, C_5$ & $\mathbf{1}$, $\mathbf{1}$, $\mathbf{1}''$ & $1,3,1$ & $1$ & --- \\ \hline	 	
$\mathcal{A}_6$ & $W_1, C_6$ & $\mathbf{1}'$, $\mathbf{1}'$, $\mathbf{1}$ & $1,3,1$ & $1$ & --- \\ \hline	
$\mathcal{A}_7$ & $W_1, C_7$ & $\mathbf{1}''$, $\mathbf{1}''$, $\mathbf{1}$ & $1,3,1$ & $1$ & --- \\ \hline	
$\mathcal{A}_8$ & $W_1, C_8$ & $\mathbf{1}''$, $\mathbf{1}''$, $\mathbf{1}'$ & $1,3,1$ & $1$ & --- \\ \hline
$\mathcal{A}_9$ & $W_1, C_9$ & $\mathbf{1}'$, $\mathbf{1}'$, $\mathbf{1}''$ & $1,3,1$ & $1$ & --- \\ \hline
$\mathcal{A}_{10}$ & $W_1, C_{10}$ & $\mathbf{1}$, $\mathbf{1}''$, $\mathbf{1}'$ & $1,1,1$ & $1$ & --- \\ \hline\hline

$\mathcal{B}_{1}(\mathcal{C}_{1})[\mathcal{D}_{1}]$ & $S_1(S_2)[S_3], C_1$ & $\mathbf{1}$, $\mathbf{1}$, $\mathbf{1}$ & $0(3)[1],2(5)[3],4(7)[5]$ & $2(-1)[1]$ & $0(1)[1]$\\ \hline			
$\mathcal{B}_{2}(\mathcal{C}_{2})[\mathcal{D}_{2}]$ & $S_1(S_2)[S_3], C_2$ & $\mathbf{1}'$, $\mathbf{1}'$, $\mathbf{1}'$ & $0(3)[1],2(5)[3],4(7)[5]$ & $2(-1)[1]$ & $0(1)[1]$\\ \hline	
 $\mathcal{B}_{3}(\mathcal{C}_{3})[\mathcal{D}_{3}]$& $S_1(S_2)[S_3], C_3$ & $\mathbf{1}''$, $\mathbf{1}''$, $\mathbf{1}''$ & $0(3)[1],2(5)[3],4(7)[5]$ & $2(-1)[1]$ & $0(1)[1]$\\ \hline	
$\mathcal{B}_{4}(\mathcal{C}_{4})[\mathcal{D}_{4}]$ & $S_1(S_2)[S_3], C_4$ & $\mathbf{1}$, $\mathbf{1}$, $\mathbf{1}'$ & $0(3)[1],2(5)[3],0(3)[1]$ & $2(-1)[1]$ & $0(1)[1]$ \\ \hline	
$\mathcal{B}_{5}(\mathcal{C}_{5})[\mathcal{D}_{5}]$ & $S_1(S_2)[S_3],C_5$ & $\mathbf{1}$, $\mathbf{1}$, $\mathbf{1}''$ & $0(3)[1],2(5)[3],0(3)[1]$ & $2(-1)[1]$ & $0(1)[1]$\\ \hline	 	
$\mathcal{B}_{6}(\mathcal{C}_{6})[\mathcal{D}_{6}]$ & $S_1(S_2)[S_3], C_6$ & $\mathbf{1}'$, $\mathbf{1}'$, $\mathbf{1}$ & $0(3)[1],2(5)[3],0(3)[1]$ & $2(-1)[1]$ & $0(1)[1]$\\ \hline	
 $\mathcal{B}_{7}(\mathcal{C}_{7})[\mathcal{D}_{7}]$ & $S_1(S_2)[S_3], C_7$ & $\mathbf{1}'$, $\mathbf{1}'$, $\mathbf{1}''$ & $0(3)[1],2(5)[3],0(3)[1]$ & $2(-1)[1]$ & $0(1)[1]$\\ \hline
$\mathcal{B}_{8}(\mathcal{C}_{8})[\mathcal{D}_{8}]$ & $S_1(S_2)[S_3], C_8$ & $\mathbf{1}''$, $\mathbf{1}''$, $\mathbf{1}$ & $0(3)[1],2(5)[3],0(3)[1]$ & $2(-1)[1]$ & $0(1)[1]$\\ \hline
$\mathcal{B}_{9}(\mathcal{C}_{9})[\mathcal{D}_{9}]$ & $S_1(S_2)[S_3],\, C_9$ & $\mathbf{1}''$, $\mathbf{1}''$, $\mathbf{1}'$ & $0(3)[1],2(5)[3],0(3)[1]$ & $2(-1)[1]$ & $0(1)[1]$\\ \hline
$\mathcal{B}_{10}(\mathcal{C}_{10})[\mathcal{D}_{10}]$& $S_1(S_2)[S_3], C_{10}$ & $\mathbf{1}$, $\mathbf{1}''$, $\mathbf{1}'$ & $0(3)[1],0(3)[1],0(3)[1]$ & $2(-1)[1]$ & $0(1)[1]$\\ \hline\hline	      	      	
\end{tabular}}
\caption{\label{tab:poss_models}Summary of the minimal neutrino mass models with the $A_4$ modular symmetry. Notice that the neutrino masses are described by the Weinberg operator in $\mathcal{A}_i$, and the models $\mathcal{B}_i$, $\mathcal{C}_i$ and $\mathcal{D}_i$ $(i=1,\ldots,10)$ are based on the type I seesaw mechanism and they differ in the Dirac neutrino Yukawa coupling $f_N\left(Y\right)$ and the right-handed neutrino mass term $f_M\left(Y\right)$. }
\end{table*}

\section{\label{sec:numerical} Phenomenological predictions }

In the following, we shall investigate whether the models summarized in table~\ref{tab:poss_models} can be compatible with the experimental data for certain values of the free parameters. It is notable that some phases are physically irrelevant and can be absorbed by field redefinition. For example, the coupling constants $\alpha$, $\beta$, $\gamma$ and $\gamma_1$ in the charged lepton mass matrix can be taken to be positive and real by rephasing the right-handed charged lepton superfields $E^c_{1,2,3}$, while it is impossible to remove the phase of $\gamma_2$ simultaneously. As a consequence, the charged lepton mass matrix will depend on four real parameters $\beta/\alpha$, $\gamma_1/\alpha$, $|\gamma_2/\alpha|$, $\arg{(\gamma_2/\alpha)}$ for the models $\mathcal{C}_{1,2,3}$ and only two real parameters $\beta/\alpha$, $\gamma/\alpha$ for the remaining models $\mathcal{C}_{i}$ ($i=4,\ldots,10$) besides the energy scale $\alpha v_d$. As regards the neutrino sector, each element of the light neutrino mass matrix is a modular form which is a function of the complex modulus $\tau$. If the neutrino masses originate from the Weinberg operator, the effective neutrino mass matrix is determined by $\tau$ and the overall factor $v^2_u/\Lambda$. If the neutrino masses arise from the type I seesaw mechanism, the light neutrino mass matrix depends on two real parameters $|g_2/g_1|$, $\arg{(g_2/g_1)}$ and the mass scale $g^2_1v^2_u/\Lambda$ (or $g^2v^2_u/\Lambda$) which controls the absolute neutrino masses, as can seen from table~\ref{tab:neutrino}. We summarize the free parameters of each model in table~\ref{tab:freeparams}.

\begin{table}[t!]
\centering
\resizebox{1.05\textwidth}{!}{
\begin{tabular}{|c|c|c|} \hline\hline
Models & model parameters & overall scales \\ \hline
 $\mathcal{A}_{1} \sim\mathcal{A}_{3}$ & $\Re\tau, \Im\tau, \beta/\alpha, \gamma_1/\alpha, |\gamma_2/\alpha|, \arg{(\gamma_2/\alpha)}$ & $\alpha v_d$, $v_u^2/\Lambda$  \\[+0.03in] \hline
 $\mathcal{A}_{4} \sim\mathcal{A}_{10}$ & $\Re\tau$, $\Im\tau$, $\beta/\alpha$, $\gamma/\alpha $ & $\alpha v_d$, $v_u^2/\Lambda$\\[+0.03in] \hline\hline

$\mathcal{B}_{1}[\mathcal{D}_{1}] \sim \mathcal{B}_{3}[\mathcal{D}_{3}]$& {$\! \begin{aligned}
& \Re\tau, \Im\tau, \beta/\alpha, \gamma_1/\alpha, |\gamma_2/\alpha|,\\[-0.04in]
& \arg{(\gamma_2/\alpha)}, |g_2/g_1|, \arg{(g_2/g_1)}
\end{aligned} $} & $\alpha v_d,\,g_1^2 v_u^2/\Lambda$   \\[+0.03in] \hline

$\mathcal{B}_{4}[\mathcal{D}_{4}] \sim \mathcal{B}_{10}[\mathcal{D}_{10}]$& $\Re\tau$, $\Im\tau$, $\beta/\alpha$, $\gamma/\alpha$, $|g_2/g_1|$, $\arg{(g_2/g_1)}$ & $\alpha v_d$, $g_1^2v_u^2/\Lambda$ \\[+0.03in] \hline	
 		
$\mathcal{C}_{1} \sim \mathcal{C}_{3}$ & $\Re\tau$, $\Im\tau$, $\beta/\alpha$, $\gamma_1/\alpha$, $|\gamma_2/\alpha|$, $\arg{(\gamma_2/\alpha)}$ & $\alpha v_d,\,g^2v_u^2/\Lambda$\\ [+0.03in] \hline
$\mathcal{C}_{4} \sim \mathcal{C}_{10}$ & $\Re\tau$, $\Im\tau$, $\beta/\alpha$, $\gamma/\alpha$ & $\alpha v_d,\,g^2v_u^2/\Lambda$\\[+0.03in] \hline\hline    	
\end{tabular} }
\caption{The independent free parameters of the models in table~\ref{tab:poss_models}, where the physically irrelevant phases have been absorbed into the fields such that the input parameters take real and positive values.\label{tab:freeparams} }
\end{table}

The values of the free parameters in each model given in table~\ref{tab:freeparams} (but not the overall scales) are determined by the six dimensionless observable quantities:
\begin{equation}
\sin^2\theta_{12}, \sin^2\theta_{13}, \sin^2\theta_{23}, \Delta m^2_{21}/\Delta m^2_{3\ell}, m_e/m_\mu, m_\mu/m_\tau\,,
\end{equation}
where $\Delta m^2_{21}=m_2^2- m_1^2$, $\Delta m^2_{3\ell}=m_3^2-m_1^2>0$ for NO and $\Delta m^2_{3\ell}=m_3^2-m_2^2 < 0$ for IO~\cite{Esteban:2018azc}.

In order to exploring the parameter space fully and efficiently, we use the popular scan tool \texttt{MultiNest}~\cite{Feroz:2007kg,Feroz:2008xx}. This has advantages over traditional approaches,
for instance, $\chi^2$ optimization by a grid or random sample, using
pre-determined ranges and step sizes for each parameter, where the number of points required scales as $k^N$, where $N$ is the dimensions of the parameter space and $k$ is the number of points chosen for each parameter.
In such a traditional approach, as $N$ increases, the number of points in parameter space rises exponentially so much so that this approach becomes highly inefficient. Also, key information for narrow ''wedges'' region of parameter space can be missed in such an approach.

In the \texttt{MultiNest} approach followed here,
in order to quantitatively measure how well the models can describe the experimental data, we use a $\chi^2$ function defined in the usual way
to serve as a test-statistic for the goodness-of-fit. The central values and $1\sigma$ errors of the oscillation parameters are taken from~\cite{Esteban:2018azc}, and the charged lepton mass ratios $m_e/m_{\mu}$ and $m_{\mu}/m_{\tau}$ are from~\cite{Feruglio:2014jla,Ross:2007az}. Since the indication of a preferred value of the Dirac CP violating phase $\delta_{CP}$ coming from global data analyses is rather weak~\cite{Esteban:2018azc}, we do not include  the contribution from $\delta_{CP}$ to the $\chi^2$ function.
By scanning the parameter space, we find the minimum $\chi^2$ values, and hence determine the best fit values of the free dimensionless parameters. Finally, to determine the overall scale factors, we use the two quantities which have absolute magnitude, i.e.~$m_e$ and $\Delta m^2_{21}$, which are the best measured dimensional quantities in the charged lepton and neutrino sectors. We randomly vary the free parameters space in the following regions,
\begin{eqnarray}
\nonumber&&\quad~\arg{(\gamma_2/\alpha)}, \arg{(g_2/g_1)}\in [0,2\pi)\,, \\
&&\beta/\alpha, \gamma/\alpha, \gamma_1/\alpha, |\gamma_2/\alpha|, |g_2/g_1|\in [0,10^4]\,.
\end{eqnarray}
The complex modulus $\tau$ is restricted to lie in the fundamental domain, since the underlying theory has the modular symmetry $\overline{\Gamma}$, and consequently vacua related by modular transformations are physically equivalent~\cite{Novichkov:2018ovf}. Moreover, under the transformation
\begin{equation}
\label{eq:conjParams}\tau \rightarrow -\tau^{\star},\quad g_i \to g_i^{\star}\,,
\end{equation}
the mass matrices become complex conjugated, hence
the lepton masses and mixing angles are unchanged while the signs of both Dirac and Majorana CP phases are reversed~\cite{Novichkov:2018ovf}. As a consequence, it is sufficient to limit the range $\Re\tau>0$ in the numerical analysis. So in practice, we restrict
$\tau$ to be in the right-hand part of the fundamental region, as follows:  $\Re\tau \in $ $[0, 0.5]$, $\Im\tau >0$, $|\tau|>1$.
The predictions of the mixing parameters in the left-hand part of the fundamental region $\Re\tau\in [-0.5, 0]$ can simply be obtained by shifting the overall signs of the Dirac as well as Majorana CP phases. Hence all the numerical results as well as figures given in the following come in pairs with opposite CP violating phases. We list the final numerical results in the following subsection.

\subsection{Numerical results of the models}
\begin{table}[h!]
\centering	
\begin{tabular}{|c|c|c||c|c|c||c|c|c||c|c|c|} \hline\hline
Models & NO & IO & Models & NO & IO  & Models & NO & IO & Models & NO & IO \\ \hline

$\mathcal{A}_{1}$&\ding{56}&\ding{56}&$\mathcal{B}_{1}$&\ding{52}&\ding{52}&$\mathcal{C}_{1}$&\ding{56}&\ding{56}&$\mathcal{D}_{1}$&\ding{52}&\ding{52}\\ \hline
$\mathcal{A}_{2}$&\ding{56}&\ding{56}&$\mathcal{B}_{2}$&\ding{52}&\ding{52}&$\mathcal{C}_{2}$&\ding{56}&\ding{56}&$\mathcal{D}_{2}$&\ding{52}&\ding{52}\\ \hline
$\mathcal{A}_{3}$&\ding{56}&\ding{56}&$\mathcal{B}_{3}$&\ding{52}&\ding{52}&$\mathcal{C}_{3}$&\ding{56}&\ding{56}&$\mathcal{D}_{3}$&\ding{52}&\ding{52}\\ \hline
$\mathcal{A}_{4}$&\ding{56}&\ding{56}&$\mathcal{B}_{4}$&\ding{56}&\ding{56}&$\mathcal{C}_{4}$&\ding{56}&\ding{56}&$\mathcal{D}_{4}$&\ding{56}&\ding{52}\\ \hline
$\mathcal{A}_{5}$&\ding{56}&\ding{56}&$\mathcal{B}_{5}$&\ding{56}&\ding{56}&$\mathcal{C}_{5}$&\ding{56}&\ding{56}&$\mathcal{D}_{5}$&\ding{52}&\ding{56}\\ \hline
$\mathcal{A}_{6}$&\ding{56}&\ding{56}&$\mathcal{B}_{6}$&\ding{56}&\ding{52}&$\mathcal{C}_{6}$&\ding{56}&\ding{56}&$\mathcal{D}_{6}$&\ding{52}&\ding{56}\\ \hline
$\mathcal{A}_{7}$&\ding{56}&\ding{56}&$\mathcal{B}_{7}$&\ding{56}&\ding{56}&$\mathcal{C}_{7}$&\ding{56}&\ding{56}&$\mathcal{D}_{7}$&\ding{52}&\ding{52}\\ \hline
$\mathcal{A}_{8}$&\ding{56}&\ding{56}&$\mathcal{B}_{8}$&\ding{56}&\ding{56}&$\mathcal{C}_{8}$&\ding{56}&\ding{56}&$\mathcal{D}_{8}$&\ding{52}&\ding{52}\\ \hline
$\mathcal{A}_{9}$&\ding{56}&\ding{56}&$\mathcal{B}_{9}$&\ding{52}&\ding{52}&$\mathcal{C}_{9}$&\ding{56}&\ding{56}&$\mathcal{D}_{9}$&\ding{52}&\ding{52}\\ \hline
$\mathcal{A}_{10}$&\ding{56}&\ding{56}&$\mathcal{B}_{10}$&\ding{52}&\ding{52}&$\mathcal{C}_{10}$&\ding{56}&\ding{56}&$\mathcal{D}_{10}$&\ding{52}&\ding{52}\\ \hline\hline
\end{tabular}
\caption{The summary of numerical results of all models for NO and IO orderings. ``\ding{52}'' signifies the models whose best-fit values fall in the $3\sigma$ range of the global fits of the experimental results~\cite{Esteban:2018azc}. In contrast, ``\ding{56}'' means that the best-fit values of the models exceed the $3\sigma$ range of the experimental data~\cite{Esteban:2018azc}. It can be seen that the models $\mathcal{A}_{1} \sim \mathcal{A}_{10}$ and $\mathcal{C}_{1} \sim \mathcal{C}_{10}$ are not consistent with the experimental data. \label{tab:poss_models_numerical_result}	}
\end{table}

We have extensively scanned over the parameter space of for each model. The results of the numerical analysis are summarized in table~\ref{tab:poss_models_numerical_result}. Henceforth we focus on the details of the numerical results of the some of these models whose predictions can lie in the $3\sigma$ range of the experimental data~\cite{Esteban:2018azc},
which are denoted by "\ding{52}". Our main interest is the case of NO ordering, preferred by the latest global fits, in particular those models containing as few parameters as possible. Thus we provide a detailed numerical analysis of the models
$\mathcal{B}_{9},\,\mathcal{B}_{10},\,\mathcal{D}_{5} \sim \mathcal{D}_{10}$ with eight parameters giving NO ordering (where $\mathcal{D}_{10}$ is the original model presented in~\cite{Feruglio:2017spp} and the other examples are new cases discussed here for the first time). For the case of IO ordering, we just give one example: model $\mathcal{B}_{10}$. Later we also present detailed numerical results for the successful cases
$\mathcal{B}_{1,2,3}$ which contain two more free parameters.

\begin{figure}[htpb]
\centering
\includegraphics[width=6.5in]{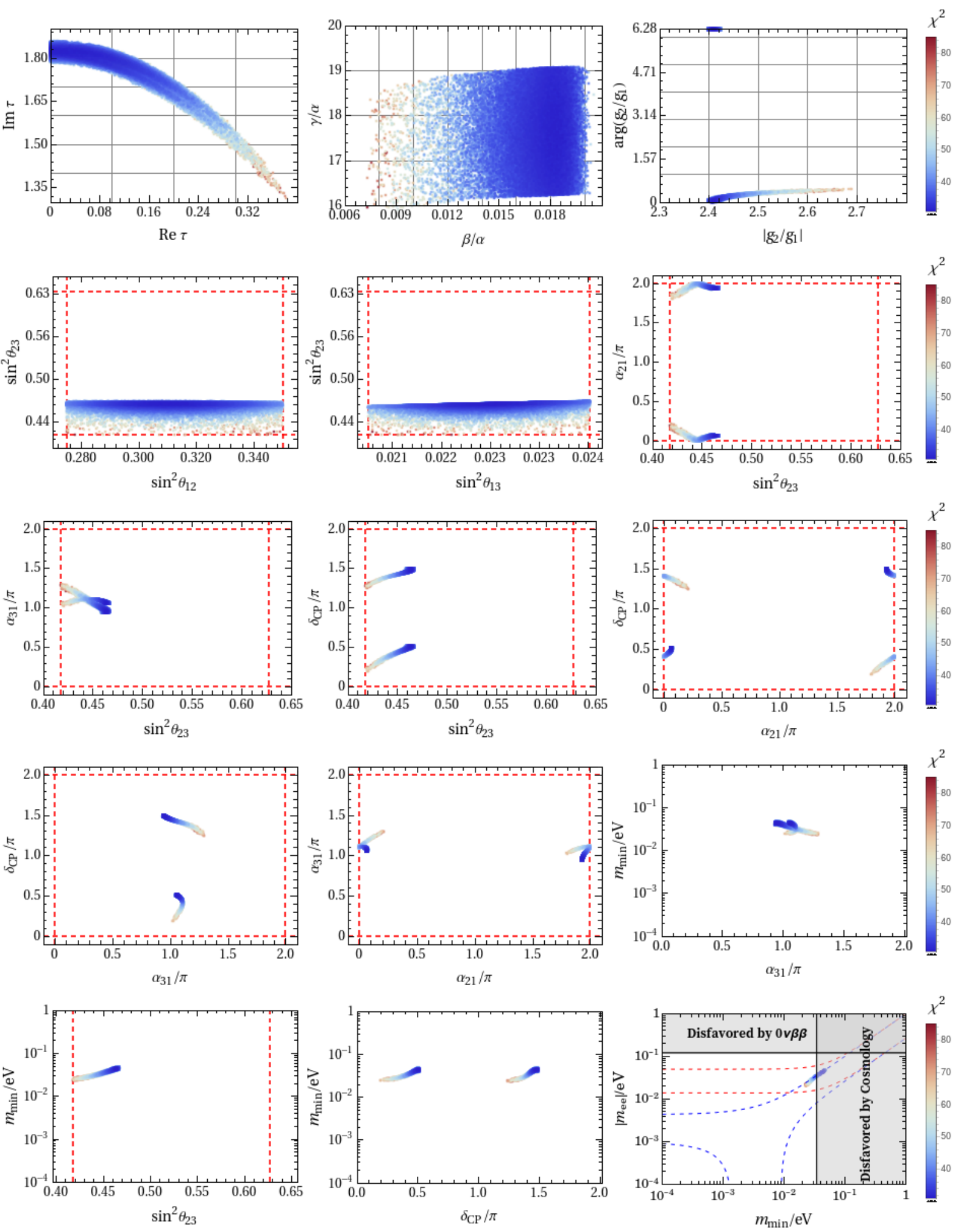}
\caption{The predictions for the correlations among the input free parameters, neutrino mixing angles, CP violation phases and neutrino masses in the model $\mathcal{B}_{9}$ with NO. The $3\sigma$ bounds of the mixing angles are shown by vertical red dashed lines~\cite{Esteban:2018azc}.
Since $\delta_{CP}$ is less constrained, we allow the regions to be in the range $0 \sim 2\pi$, and similarly for $\alpha_{21}$ and $\alpha_{31}$.
The last panel of $|m_{ee}|$ versus $m_{\text{min}}$ indicates large tightly constrained values for both these neutrino mass observables.}
\label{fig:B9_NH}
\end{figure}

\begin{figure}[htpb]
\centering
\includegraphics[width=6.5in]{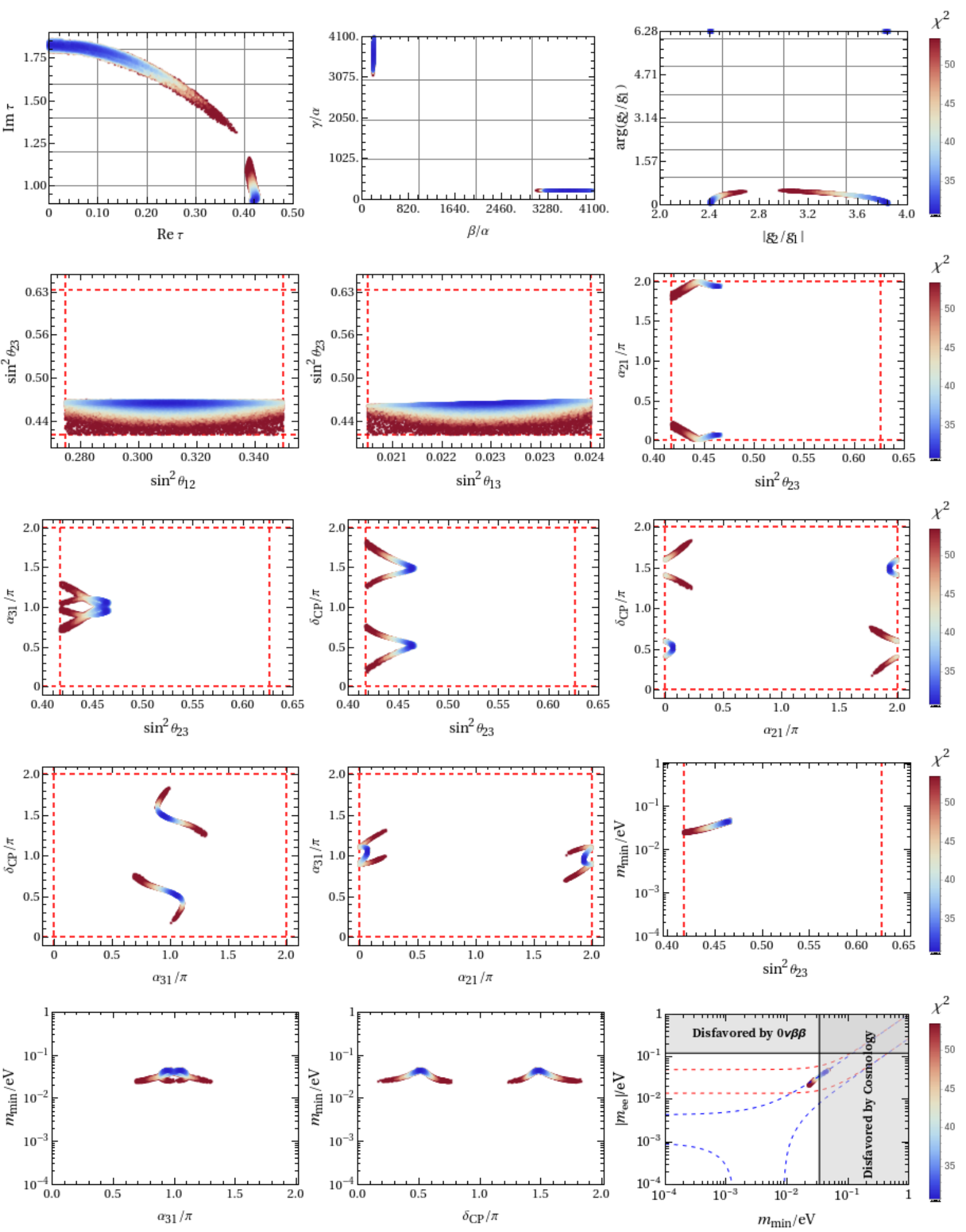}
\caption{The predictions for the correlations among the input free parameters, neutrino mixing angles, CP violation phases and neutrino masses in the model $\mathcal{B}_{10}$ with NO. Here we adopt the same conventions as figure~\ref{fig:B9_NH}.}
\label{fig:B10_NH}
\end{figure}

\begin{figure}[htpb]
\centering
\includegraphics[width=6.5in]{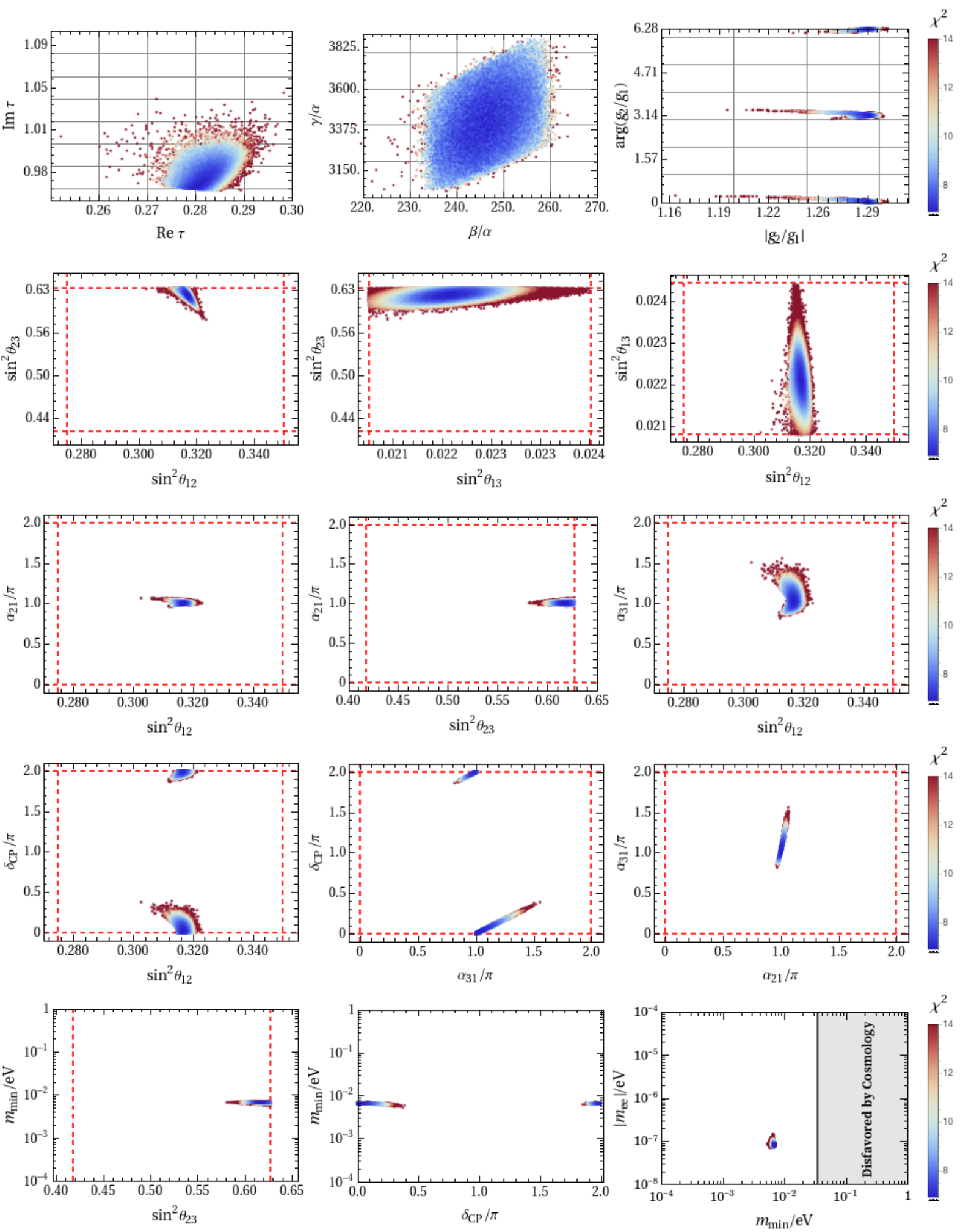}
\caption{The predictions for the correlations among the input free parameters,
neutrino mixing angles, CP violation phases and neutrino masses in the model $\mathcal{D}_{5}$ with NO. Here we adopt the same conventions as figure~\ref{fig:B9_NH}.}
\label{fig:D5_NH}
\end{figure}

\begin{figure}[htpb]
\centering
\includegraphics[width=6.5in]{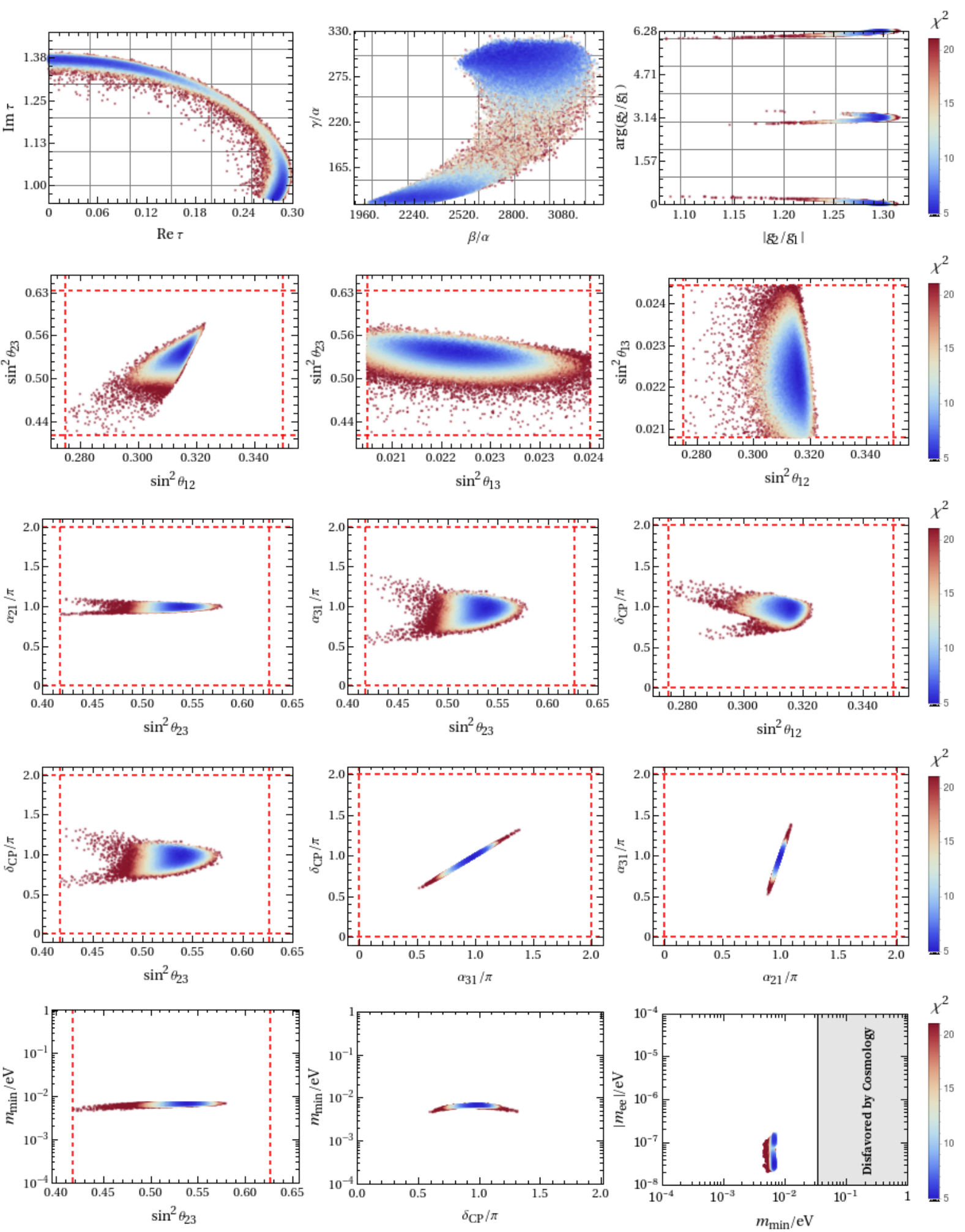}
\caption{The predictions for the correlations among the input free parameters,
neutrino mixing angles, CP violation phases and neutrino masses in the model $\mathcal{D}_{6}$ with NO. Here we adopt the same conventions as figure~\ref{fig:B9_NH}.}
\label{fig:D6_NH}
\end{figure}

\begin{figure}[htpb]
\centering
\includegraphics[width=6.5in]{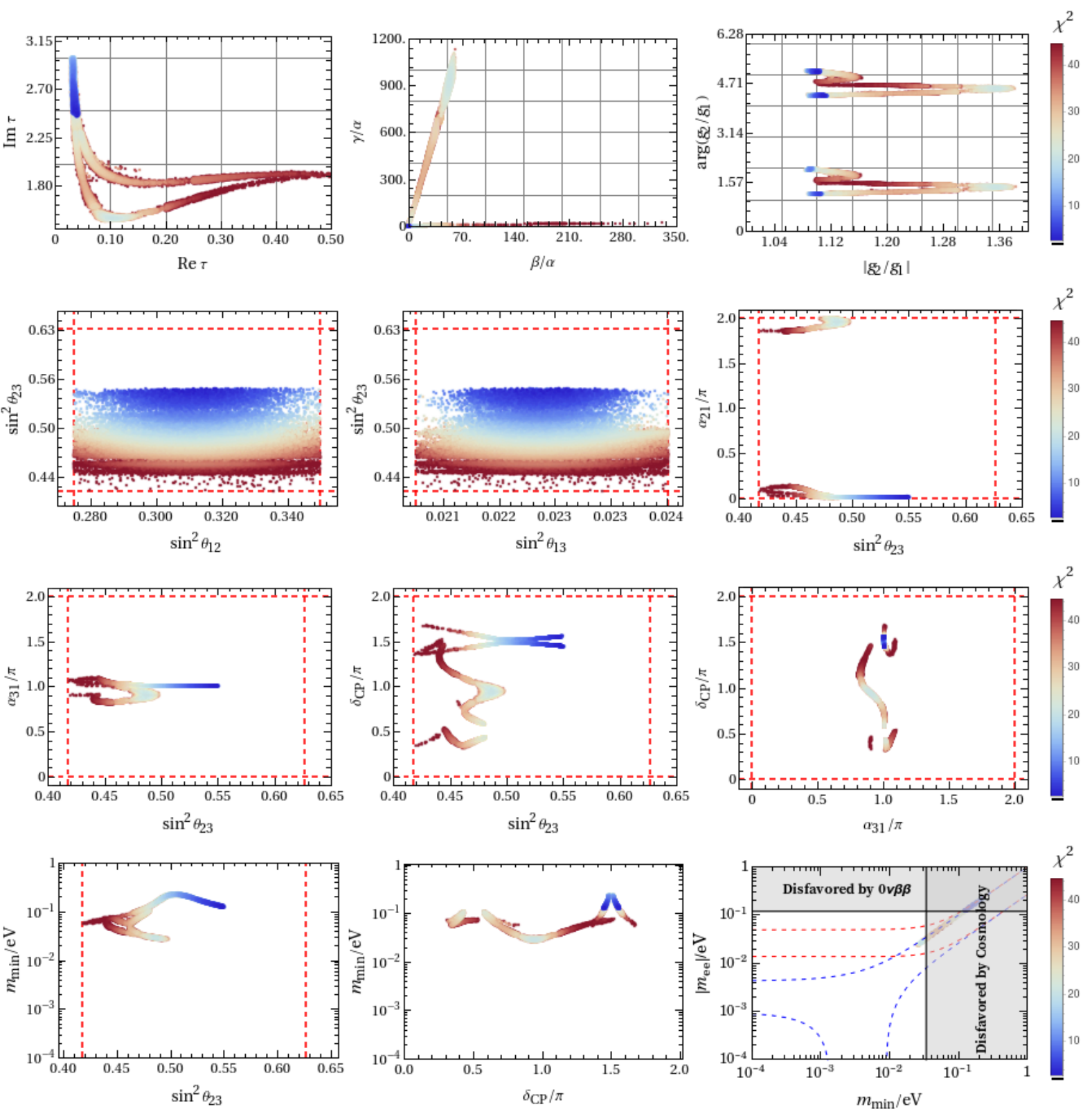}
\caption{The predictions for the correlations among the input free parameters,
neutrino mixing angles, CP violation phases and neutrino masses in the model $\mathcal{D}_{7}$ with NO. Here we adopt the same conventions as figure~\ref{fig:B9_NH}.}
\label{fig:D7_NH}
\end{figure}

\begin{figure}[htpb]
\centering
\includegraphics[width=6.5in]{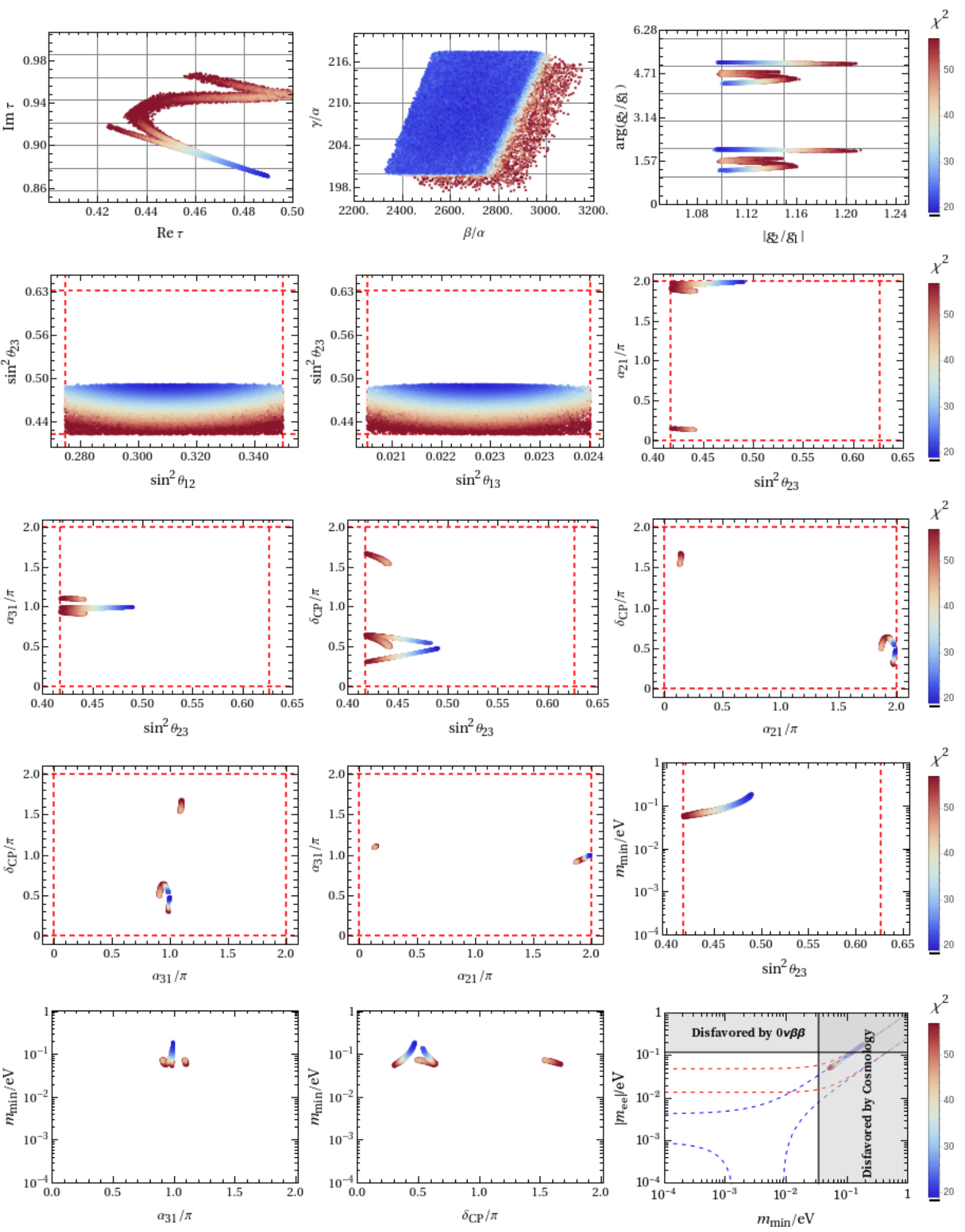}
\caption{The predictions for the correlations among the input free parameters
neutrino mixing angles, CP violation phases and neutrino masses in the model $\mathcal{D}_{8}$ with NO. Here we adopt the same conventions as figure~\ref{fig:B9_NH}.}
\label{fig:D8_NH}
\end{figure}

\begin{figure}[htpb]
\centering
\includegraphics[width=6.5in]{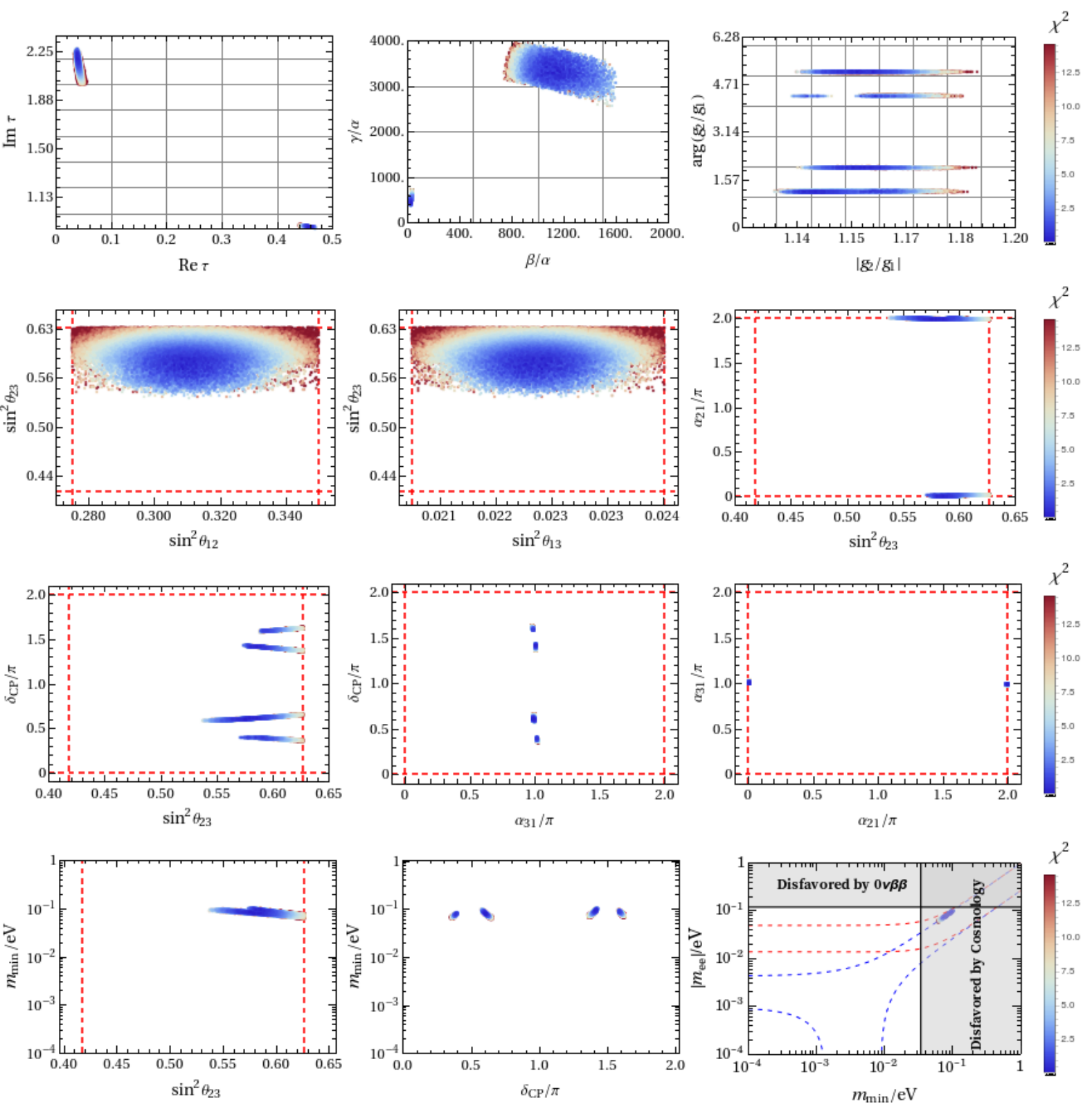}
\caption{The predictions for the correlations among the input free parameters
neutrino mixing angles, CP violation phases and neutrino masses in the model $\mathcal{D}_{9}$ with NO. Here we adopt the same conventions as figure~\ref{fig:B9_NH}.}
\label{fig:D9_NH}
\end{figure}

\begin{figure}[htpb]
\centering
\includegraphics[width=6.5in]{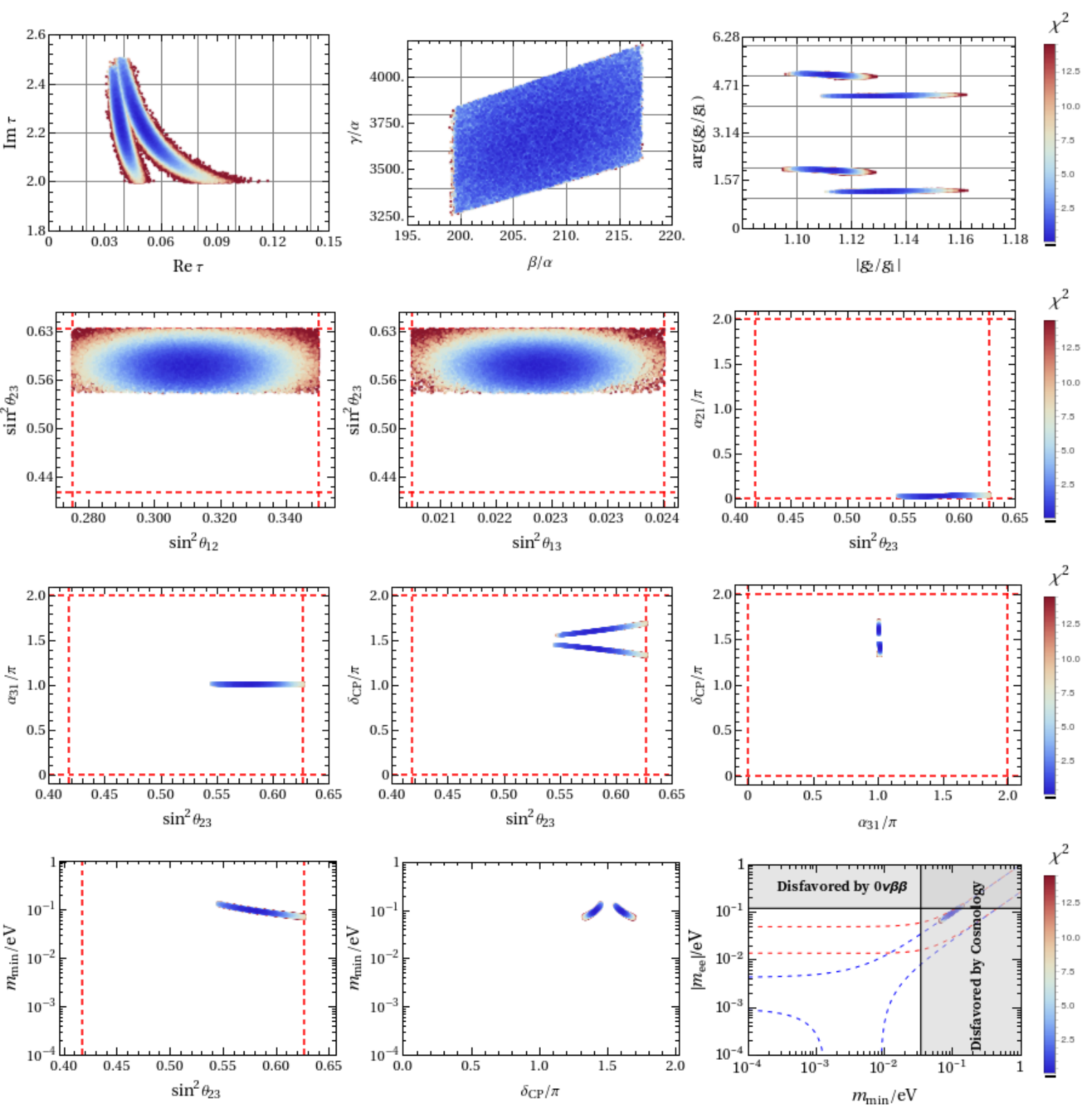}
\caption{The predictions for the correlations among the input free parameters, neutrino mixing angles, CP violation phases and neutrino masses in the model $\mathcal{D}_{10}$ with NO. Here we adopt the same conventions as figure~\ref{fig:B9_NH}.}
\label{fig:D10_NH}
\end{figure}

\begin{figure}[htpb]
\centering
\includegraphics[width=6.5in]{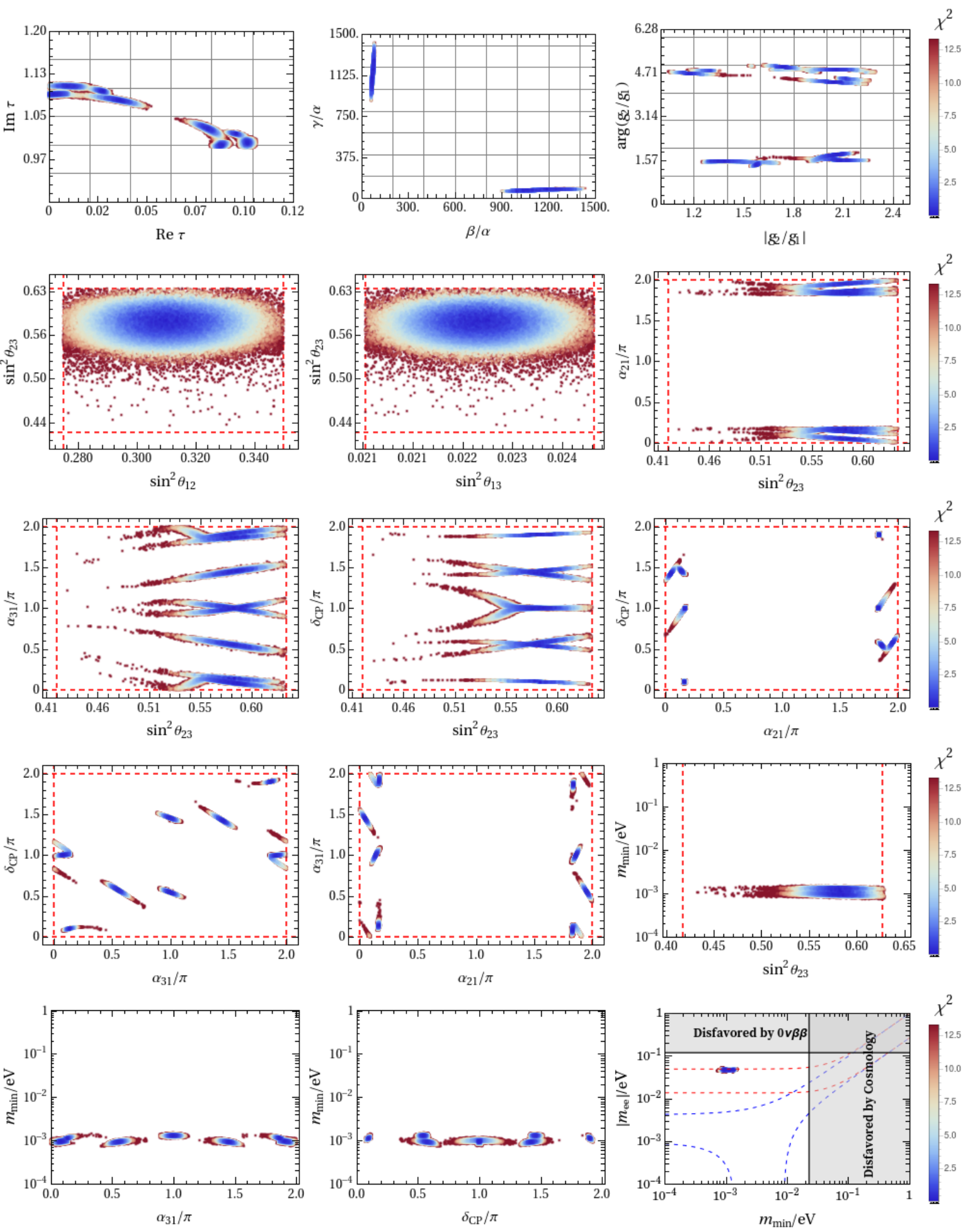}
\caption{The predictions for the correlations among the input free parameters, neutrino mixing angles, CP violation phases and neutrino masses in the model $\mathcal{B}_{10}$ with IO. Here we adopt the same conventions as figure~\ref{fig:B9_NH}.}
\label{fig:B10_IH}
\end{figure}

The results of the numerical analysis are summarized in tables \ref{tab:B9B10_NH}-\ref{tab:B10_IH}. In particular we highlight the new cases $\mathcal{D}_{7}$ and $\mathcal{D}_{9}$ which have a very small
$\chi^2$ and predict $\delta_{CP}/\pi \approx 1.42-1.45$. We display some interesting correlations of the parameters and observables in these models in figures~\ref{fig:B9_NH}-\ref{fig:B10_IH}, where the colour of the points in these figures indicates the corresponding $\chi^2$ value. Note that many of these figures show very tightly constrained regions of observable parameters. For models $\mathcal{B}_{1,2,3}$ and $\mathcal{D}_{1,2,3}$ with two more parameters (which can be see from table~\ref{tab:freeparams}), we only report the predictions for the observables at the best-fit point, with the results summarized in table~\ref{tab:B1B2B3_D1D2D3}. The allowed regions of the input parameters and observables are determined by requiring all the lepton mixing angles and the squared mass splittings
$\Delta m^2_{21}$ and $\Delta m^2_{31}$ ($\Delta m^2_{32}$) within the $3\sigma$ intervals~\cite{Esteban:2018azc}.

Most of these models $\mathcal{B}_{9}$, $\mathcal{B}_{10}$, $\mathcal{D}_{5} \sim \mathcal{D}_{10}$ (apart from $\mathcal{D}_{5}$ and $\mathcal{D}_{6}$)
predict large (but allowed) neutrino masses and observable neutrinoless double beta decay. The latest Planck result on the neutrino mass sum is $\sum_{i}m_{i}<0.12~\mathrm{eV}-0.60~\mathrm{eV}$~\cite{Aghanim:2018eyx}.
Since the upper bound of the neutrino mass sum sensitively depends on the cosmological model and the choice of other experimental data, we display the full range $0.12~\mathrm{eV}-0.60~\mathrm{eV}$ as ``disfavoured by cosmology'' in the figures. Our predictions for neutrino masses could also be probed in next generation neutrinoless double beta decay experiments which is the only feasible experiment having the potential of establishing Majorana nature of neutrinos. The measurement of neutrinoless double beta decay could provide unique information on the neutrino mass spectrum, Majorana phases and the absolute scale of neutrino masses.
The decay amplitude is proportional to the effective Majorana mass $m_{ee}$ with the absolute value~\cite{Tanabashi:2018oca},
\begin{equation}
|m_{ee}|=\left|m_1c^2_{12}c^2_{13}+m_2s^2_{12}c^2_{13}e^{i\alpha_{21}}+m_3s^2_{13}e^{i(\alpha_{31}-2\delta_{CP})}\right|\,.
\end{equation}
The neutrinoless double beta decay experiments can provide valuable information on the neutrino mass spectrum and constrain the Majorana phases. Most of the above models predict neutrino masses in the ``cosmologically disfavoured region'' and observable neutrinoless double beta decay, which can be tested in forthcoming experiments, with the exception of $\mathcal{D}_{5}$ and $\mathcal{D}_{6}$ which however predict tiny neutrinoless double beta decay, deep into the NO ``hole'', together with
small Dirac CP violation.

\begin{table}[htpb]
\centering
\renewcommand{\arraystretch}{1.2}
\resizebox{1.1\textwidth}{!}{\begin{tabular}{|c|cc|cc|} \hline\hline
& \multicolumn{2}{c|}{ Model $\mathcal{B}_{9}$} & \multicolumn{2}{c|}{Model $\mathcal{B}_{10}$} \\ \cline{2-5}
&  \multicolumn{2}{c|}{NO}                &  \multicolumn{2}{c|}{NO} \\ \cline{2-5}
& Best-fit  &   Allowed regions & Best-fit  &  Allowed regions \\  \hline

$\Re\langle\tau\rangle$  & 0.0003 & $[0, 0.368]$  & 0.0129 & $[0, 0.431]$\\
$\Im\langle\tau\rangle$  & 1.824 & $[1.351, 1.856]$    & 1.824 &  $[0.91, 1.16]\cup[1.31,1.86] $\\
$\beta/\alpha$ & 0.018 & $[0.008, 0.020]$  & 205.720 & $[192.39,215]\cup[3054.25 4093.49]$\\
$\gamma/\alpha$  & 17.560 & $[16.046, 19.063]$  & 3612.07 & $[192.4,215]\cup[3066.47, 4092.98]$\\
$|g_2/g_1|$   & 2.410 & $[2.399, 2.701]$  & 2.410 & $[2.398,2.71]\cup[2.95,3.86]$ \\
$\arg{(g_2/g_1)}$  & 0.030 & $ [0,0.47]\cup[6.25, 2\pi]$ & 6.267 & $[0, 0.49] \cup [6.23,2\pi]$ \\
$\alpha v_d$/MeV   & 106.523 & ---     & 0.5179 & ---\\
$(g^2_1v_u^2/\Lambda)$/eV   & 0.011 & ---  & 0.0111 & ---  \\ \hline
$m_e/m_{\mu}$    & 0.0048 & $[0.0046, 0.0050]$    & 0.0048 & $[0.0046, 0.0050]$ \\
$m_{\mu}/m_{\tau}$  & 0.0564 & $[0.0520, 0.0610]$  & 0.0564 & $[0.0520, 0.0610]$ \\ \hline
$\sin^2\theta_{12}$  & 0.3096 & $[0.2750, 0.3500]$   & 0.3096 & $[0.2750, 0.3500]$ \\
$\sin^2\theta_{13}$  & 0.02263 & $[0.02046, 0.02439]$  & 0.0226  & $[0.02045, 0.02439]$\\
$\sin^2 \theta_{23}$  & 0.4637 & $[0.4180, 0.4674]$ & 0.4638 & $[0.4180, 0.4676]$ \\
$\delta_{CP}/\pi$    & 0.510 & $[0.17,0.51]\cup [1.24,1.5]$   & 1.486 &  $[0.19,0.77]\cup [1.23,1.84]$\\
$\alpha_{21}/\pi$   & 0.068 & $ [0,0.23]\cup [1.78, 2]$     & 0.068 & $[0,0.23]\cup [1.77,2]$\\
$\alpha_{31}/\pi$  & 1.056 & $[0.937, 1.270]$   & 0.948 & $[0.683, 1.311]$\\ \hline
$m_1$/eV   & 0.0430 & $[0.0228, 0.0476]$   & 0.0430 & $[0.0225, 0.0478]$\\
$m_2$/eV   & 0.0438 & $[0.0244, 0.0484]$   & 0.0439 & $[0.0241, 0.0485]$ \\
$m_3$/eV   & 0.0661 & $[0.0525, 0.0716]$  & 0.0661 & $[0.0524, 0.0716]$ \\
$\sum_i m_i$/eV & 0.1529 & $[0.0997, 0.1676]$   & 0.1530 & $[0.0991, 0.1679]$\\
$|m_{ee}|$/eV   & 0.0435 & $[0.0206, 0.0483]$   & $0.0436$ &  $[0.0202, 0.0483]$ \\ \hline
$\chi^2_{\mathrm{min}}$                         & 30.77 & ---                        & 30.72 &  --- \\ \hline\hline
\end{tabular} }
\caption{The predictions for the best-fit values and the allowed ranges of the input parameters and observables in the models $\mathcal{B}_{9}$ and $\mathcal{B}_{10}$ with NO. We would like to emphasize that the Dirac CP phase $\delta_{CP}\simeq1.49\pi$ at the conjugate best fit point $\tau\rightarrow-\tau^{*}, g_i\rightarrow g^{*}_i$ in model $\mathcal{B}_{9}$.\label{tab:B9B10_NH}}
\end{table}

\begin{table}[htpb]
\centering
\renewcommand{\arraystretch}{1.2}
\resizebox{\textwidth}{!}{
\begin{tabular}{|c|cc|cc|} \hline\hline
& \multicolumn{2}{c|}{ Model $\mathcal{D}_{5}$ } & \multicolumn{2}{c|}{ Model $\mathcal{D}_{6}$ } \\ \cline{2-5}
&  \multicolumn{2}{c|}{NO}                &  \multicolumn{2}{c|}{NO} \\ \cline{2-5}
& Best-fit  &  Allowed regions & Best-fit  &  Allowed regions \\  \hline

$\Re\langle\tau\rangle$  & 0.280 & $[0.248, 0.300]$   & 0.279 & $[0.0, 0.300]$\\
$\Im\langle\tau\rangle$   & 0.960 & $[0.957, 1.056]$   & 0.960 & $[0.957, 1.394] $\\
$\beta/\alpha$  & 244.708 & $[216.359, 266.608]$  & 2774.426 & $[1954.8, 3290.0]$\\
$\gamma/\alpha$   & 3397.7 & $[2991.9, 3880.4]$ & 302.315 & $[118.236, 327.297]$\\
$|g_2/g_1|$   & 1.293 & $[1.143, 1.308]$   & 1.294 & $[1.034, 1.314]$ \\
\multirow{2}{*}{$\arg{(g_2/g_1)}$}    & \multirow{2}{*}{0} & $[0,0.23]\cup[3.01,3.37]$    & \multirow{2}{*}{3.142} & $[0.0,0.29]\cup[2.85,3.38]$ \\
& & $\cup[6.1, 2\pi]$ & & $\cup[5.97, 2\pi]$ \\
$\alpha v_d $/MeV & 0.4174 & ---       & 0.4177 & ---\\
$(g^2_1 v_u^2/\Lambda)$/eV    & 0.01321 & ---      & 0.01321 & ---  \\ \hline
$m_e/m_{\mu}$   & 0.0048 & $[0.0046, 0.0050]$   & 0.0048 & $[0.0046, 0.0050]$ \\
$m_{\mu}/m_{\tau}$ & 0.0569 &$[0.0520, 0.0610]$  &  0.0561  & $[0.0520, 0.0610]$ \\ \hline
$\sin^2\theta_{12}$    & 0.3169 & $[0.3011, 0.3244]$  & 0.3161 & $[0.2752, 0.3241]$ \\
$\sin^2\theta_{13}$    & 0.02189 & $[0.02045, 0.02439]$   & 0.0220 & $[0.02045, 0.02439]$\\
$\sin^2\theta_{23}$   & 0.6171 & $[0.5754, 0.627]$    & 0.5396 & $
[0.4186, 0.5845]$ \\
$\delta_{CP}/\pi$   & 0 &  $[0,0.41]\cup [1.79,2]$    & $1.0$ &  $[0.57, 1.33]$\\
$\alpha_{21}/\pi$ & 1.0 & $[0.959,1.071]$   & 1.0 & $[0.89, 1.1]$\\
$\alpha_{31}/\pi$    & 1.0 & $[0.746, 1.603]$   & 1.0 & $[0.495, 1.396]$\\ \hline
$m_1$/eV   & 0.0067 & $[0.0052, 0.0069]$   & 0.0067 & $[0.0042, 0.0069]$\\
$m_2$/eV   & 0.0109 & $[0.0100, 0.0110]$   & 0.0109 & $[0.0095, 0.0110]$ \\
$m_3$/eV   & 0.0499 & $[0.0476, 0.0520]$  & 0.0501 & $[0.0476, 0.0539]$ \\
$\sum_i m_i$/eV & 0.0675 & $[0.0628, 0.0699]$     & 0.0677 & $[0.0613, 0.0717]$\\
$|m_{ee}|$/eV   & $10^{-8}$ & $[10^{-8}, 10^{-7}]$   & $10^{-7}$ & $[10^{-8}, 10^{-7}]$ \\ \hline
$\chi^2_{\mathrm{min}}$      & 6.82 &  ---      & 4.857 &  --- \\ \hline\hline
\end{tabular} }
\caption{The predictions for the best-fit values and the allowed ranges of the input parameters and observables in the models $\mathcal{D}_{5}$ and $\mathcal{D}_{6}$ with NO. \label{tab:D5D6_NH}}
\end{table}

\begin{table}[htpb]
\centering
\renewcommand{\arraystretch}{1.2}
\resizebox{1.05\textwidth}{!}{
\begin{tabular}{|c|cc|cc|} \hline\hline
 & \multicolumn{2}{c|}{Model $\mathcal{D}_{7}$} & \multicolumn{2}{c|}{Model $\mathcal{D}_{8}$} \\ \cline{2-5}

&  \multicolumn{2}{c|}{NO}                &  \multicolumn{2}{c|}{NO} \\ \cline{2-5}
& Best-fit  &  Allowed regions & Best-fit  &  Allowed regions \\  \hline

$\Re\langle\tau\rangle$  & 0.0428 & $[0.026, 0.5]$  & 0.471 & $[0.424, 0.5]$\\

$\Im\langle\tau\rangle$   & 2.105 &  $[1.468, 3.006]$   & 0.886 &  $[0.872, 0.964]$\\

$\beta/\alpha$    & 0.473 & $[0.048, 339.73]$    & 2646.6 & $[2327.7, 3181.7]$\\

$\gamma/\alpha$     & 0.002 & $[0.002, 1140.03]$   & 208.094 & $[197.1, 217.356]$\\

$|g_2/g_1|$    & 1.154 & $[1.084, 1.385]$   & 1.113 & $[1.094, 1.212]$ \\

$\arg{(g_2/g_1)}$  & 1.964 & $[1.16,1.98]\cup[4.3, 5.12]$    & 1.227 & $[1.19,1.98]\cup[4.34, 5.12]$ \\
$\alpha v_d $/MeV    & 1702.3 & ---     & 0.368 & ---\\
$(g^2_1 v_u^2/\Lambda)$/eV   & 0.0405 & ---       & 0.036 & ---  \\ \hline
$m_e/m_{\mu}$     & 0.0048 & $[0.0046, 0.0050]$     & 0.0048 & $[0.0046, 0.0050]$ \\

$m_{\mu}/m_{\tau}$    & 0.0565 &$[0.0520, 0.0610]$   & 0.0565 & $[0.0520, 0.0610]$ \\ \hline
$\sin^2\theta_{12}$  & 0.3100 & $[0.2750, 0.3500]$  & 0.3105 & $[0.2750, 0.3500]$ \\
$\sin^2 \theta_{13}$ & 0.0224 & $[0.02045, 0.02439]$   & 0.0224 & $[0.02045, 0.02439]$\\
$\sin^2\theta_{23}$  & 0.580 & $[0.418, 0.551]$   & 0.4698 & $[0.418, 0.491]$ \\
$\delta_{CP}/\pi$    & 1.60 &  $[0.307, 1.702]$   & $1.522$ &  $[0.29,0.65]\cup[1.52,1.68]$\\
$\alpha_{21}/\pi$     & 1.99 & $[0, 0.14]\cup[1.84,2]$ & 0 & $ [0.12,0.16]\cup[1.86,2]$\\
$\alpha_{31}/\pi$    & 0.986 & $[0.806, 1.1]$    & 1.002 & $[0.898, 1.115]$\\ \hline
$m_1$/eV  & 0.0805 & $[0.0250, 0.2437]$   & 0.1003 & $[0.0505, 0.1885]$\\
$m_2$/eV  & 0.0810 & $[0.0264, 0.2438]$   & 0.1007 & $[0.0512, 0.1887]$ \\
$m_3$/eV  & 0.0949 & $[0.0537, 0.2495]$  & 0.1122 & $[0.0695, 0.1956]$ \\
$\sum_i m_i$/eV  & 0.2564 & $[0.1051, 0.7370]$   & 0.3132 & $[0.1712, 0.5729]$\\
$|m_{ee}|$/eV   & 0.0805 &  $[0.0235, 0.2438]$  & 0.1004 &  $[0.0501, 0.1887]$ \\ \hline

$\chi^2_{\mathrm{min}}$      & 0.0003 &  ---      & 27.5 &  --- \\ \hline\hline
\end{tabular} }
\caption{The predictions for the best-fit values and the allowed ranges of the input parameters and observables in the models $\mathcal{D}_{7}$ and $\mathcal{D}_{8}$ with NO. \label{tab:D7D8_NH}}
\end{table}

\begin{table}[htpb]
\centering
\renewcommand{\arraystretch}{1.2}
\resizebox{1.1\textwidth}{!}{\begin{tabular}{|c|cc|cc|} \hline\hline
& \multicolumn{2}{c|}{Model $\mathcal{D}_{9}$} & \multicolumn{2}{c|}{Model $\mathcal{D}_{10}$ } \\ \cline{2-5}

&  \multicolumn{2}{c|}{NO}                &  \multicolumn{2}{c|}{NO} \\ \cline{2-5}
& Best-fit  &  Allowed regions & Best-fit  & Allowed regions \\  \hline

$\Re\langle\tau\rangle$   & 0.0387 & $[0.033,0.056]\cup[0.44,0.469]$  & 0.0386 & $[0.0307, 0.1175]$\\

$\Im\langle\tau\rangle$  & 2.233 &  $[0.887,0.908]\cup[2.0, 2.282]$    & 2.230 &  $[1.996, 2.50] $\\

$\beta/\alpha$   & 23.195 & $[21.24,38.95]\cup[737.8,1599.9]$    & 207.908 & $[198.963, 217.263]$\\
$\gamma/\alpha$    & 410.532 & $[352.32,700]\cup[2520.65,3983.95]$  & 3673.38 & $[3254.84, 4170.84]$\\
$|g_2/g_1|$   & 1.138 & $[1.127, 1.190]$   & 1.129 & $[1.094, 1.162]$ \\
\multirow{2}{*}{$\arg{(g_2/g_1)}$} & \multirow{2}{*}{1.172} & $[1.16, 1.21]\cup[1.93,1.98]$  & \multirow{2}{*}{1.197} & $[1.17,1.25]\cup[1.76,1.95]$ \\
& & $\cup[4.3,4.35]\cup[5.07,5.13]$ & & $\cup[4.31,4.4]\cup[4.9,5.09]$ \\
$\alpha v_d $/MeV  & 4.585 & ---        & 0.512 & ---\\
$(g^2_1 v_u^2/\Lambda)$/eV  & 0.0476 & ---     & 0.0475 & ---  \\ \hline
$m_e/m_{\mu}$   & 0.0048 & $[0.0046, 0.0050]$   & 0.0048 & $[0.0046, 0.0050]$ \\
$m_{\mu}/m_{\tau}$   & 0.0565 &$[0.0520, 0.0610]$   & 0.0565 &$[0.0520, 0.0610]$ \\ \hline
$\sin^2\theta_{12}$   & 0.3098 & $[0.2750, 0.3500]$   & 0.3098 & $[0.2750, 0.3500]$ \\
$\sin^2\theta_{13}$   & 0.0224 & $[0.02045, 0.02439]$   & 0.0224 & $[0.02045, 0.02439]$\\
$\sin^2\theta_{23}$  & 0.5807 & $[0.5353, 0.6270]$  & 0.580 & $[0.5456, 0.6270]$ \\
\multirow{2}{*}{$\delta_{CP}/\pi$}  & \multirow{2}{*}{1.420} &  $[0.35,0.4]\cup[0.58,0.66]$    & \multirow{2}{*}{1.604} &  \multirow{2}{*}{$[1.33,1.45]\cup[1.55,1.7]$}\\
& & $\cup[1.36,1.43]\cup[1.58,1.63]$ & & \\
$\alpha_{21}/\pi$    & 0.006 & $[0,0.01]\cup[1.98,2]$    & 0.015 & $[0.009, 0.0373]$\\
$\alpha_{31}/\pi$   & 1.005 & $[0.978, 1.027]$  & 1.007 & $[1.003, 1.029]$\\ \hline
$m_1$/eV    & 0.0948 & $[0.0601, 0.1044]$  & 0.0946 & $[0.0658, 0.1378]$\\
$m_2$/eV   & 0.0952 & $[0.0607, 0.1048]$  & 0.0950 & $[0.0663, 0.1381]$ \\
$m_3$/eV  & 0.1073 & $[0.0765, 0.1167]$   & 0.1071 & $[0.0811, 0.1476]$ \\
$\sum_i m_i$/eV  & 0.2974 & $[0.1973, 0.3259]$    & 0.2966 & $[0.2132, 0.4234]$\\
$|m_{ee}|$/eV    & 0.0949 &  $[0.0599, 0.1045]$   & 0.0945 &  $[0.0651, 0.1379]$ \\ \hline
$\chi^2_{\mathrm{min}}$     & 0.0023 &  ---     & 0.0003 &  --- \\\hline\hline
\end{tabular} }
\caption{The predictions for the the best-fit values and the allowed ranges of the input parameters and observables in the model $\mathcal{D}_{9}$ and $\mathcal{D}_{10}$ with NO.\label{tab:D9D10_NH} }
\end{table}

\begin{table}[htpb]
\centering
\renewcommand{\arraystretch}{1.2}
\resizebox{0.7\textwidth}{!}{
\begin{tabular}{|c|cc|} \hline\hline
\multirow{2}{*}{ Model $\mathcal{B}_{10}$ }  &  \multicolumn{2}{c|}{IO} \\ \cline{2-3}
  & Best-fit & Allowed regions  \\ \hline
$\Re\langle\tau\rangle$   & 0.096 & $[0, 0.102]$\\
$\Im\langle\tau\rangle$   & 0.987 &  $[0.98, 1.049]\cup[1.052,1.109]$\\
$\beta/\alpha$  & 79.472 & $[59.68,86.37]\cup[892.67, 1446.02]$\\
$\gamma/\alpha$  & 1232.57 & $[60.97,86.34]\cup[870.16, 1443.84]$\\
$|g_2/g_1|$   & 2.093 & $[1.038, 2.453]$ \\
$\arg{(g_2/g_1)}$  & 4.715 & $[1.33, 1.83]\cup[4.29,5]$ \\
$\alpha v_d$/MeV    & 1.167 & ---\\
$(g^2_1v_u^2/\Lambda)$/eV  & 0.004 & ---  \\ \hline
$m_e/m_{\mu}$  & 0.0048 & $[0.0046, 0.0050]$ \\
$m_{\mu}/m_{\tau}$    & 0.0565 &$[0.0520, 0.0610]$ \\ \hline
$\sin^2\theta_{12}$   & 0.3100 & $[0.2750, 0.3500]$ \\
$\sin^2 \theta_{13}$  & 0.02264 & $[0.02068, 0.02463]$\\
$\sin^2 \theta_{23}$  & 0.584 & $[0.423, 0.629]$ \\
$\delta_{CP}/\pi$  & 1.458 &  $[0.068, 1.933]$\\
$\alpha_{21}/\pi$   & 0.138 & $[0,0.19]\cup[1.8,2]$\\
$\alpha_{31}/\pi$   & 0.997 & $[0, 2]$\\ \hline
$m_1$/eV   & 0.0494 & $[0.0464, 0.0526]$\\
$m_2$/eV  & 0.0501 & $[0.0472, 0.0533]$ \\
$m_3$/eV   & 0.0013 & $[0.0007, 0.0015]$ \\
$\sum_i m_i$/eV & 0.1008 & $[0.0942, 0.1074]$\\
$|m_{ee}|$/eV   & 0.0475 &  $[0.0439, 0.0516]$ \\ \hline

$\chi^2_{\mathrm{min}}$     & $10^{-7}$ &  --- \\ \hline\hline
\end{tabular} }
\caption{The predictions for the best-fit values and the allowed ranges of the input parameters and observables in the model $\mathcal{B}_{10}$ with IO.\label{tab:B10_IH} }
\end{table}

\begin{table}[hptb]
\centering
\renewcommand{\arraystretch}{1.2}
\begin{tabular}{|c|c|c|c||c|c|c|} \hline\hline
\multirow{2}{*}{ Models} & $\mathcal{B}_{1}$ & $\mathcal{B}_{2}$ & $\mathcal{B}_{3}$ & $\mathcal{D}_{1}$ & $\mathcal{D}_{2}$ & $\mathcal{D}_{3}$ \\ \cline{2-7}
   & \multicolumn{6}{c|}{ Best-fit values for NO}  \\ \hline
$\Re\langle \tau \rangle$   & 0.485   & 0.468   & 0.487  & 0.101   & 0.099    & 0.109 \\
$\Im\langle \tau \rangle$   & 1.150   & 1.222   & 1.574 & 1.250   & 1.428    & 1.359\\
$\beta/\alpha$      & 632.056 & 2610.95 & 288.448  & 111.715 & 143.544  & 253.671  \\
$\gamma_1/\alpha$    & 59.950  & 218.726 & 1177.58  & 1306.5  & 1109.25  & 21.804 \\
$|\gamma_2/\alpha|$   & 9.452   & 211.488 & 1201.62  & 796.746 & 801.233  & 3.549 \\
$\arg{(\gamma_2/\alpha)}$     & 2.871   & 3.046   & 3.523 & 4.055   & 2.487    & 4.421 \\
$|g_2/g_1|$     & 0.992   & 1.122   & 1.647 & 1.109   & 1.543    & 1.264 \\
$\arg{(g_2/g_1)}$             & 2.203   & 2.398   & 2.081 & 6.024   & 0.889    & 0.014   \\
$\alpha v_d$/MeV  & 2.374  & 0.613   & 1.499 & 1.201   & 1.597    & 6.965 \\
$(g^2_1v_u^2/\Lambda)$/eV  & 0.0109  & 0.0103  & 0.0114 & 0.0155  & 0.0122   & 0.0173  \\ \hline
$m_e/m_{\mu}$ & 0.0048  & 0.0048  & 0.0048 & 0.0048  & 0.0048   & 0.0048 \\
$m_{\mu}/m_{\tau}$ & 0.0565 & 0.0565  & 0.0565 & 0.0565  & 0.0565 & 0.0565 \\ \hline
$\sin^2\theta_{12}$ & 0.3100 & 0.3100  & 0.3100 & 0.3100  & 0.3100  & 0.3100\\
$\sin^2\theta_{13}$ & 0.02241 & 0.02241 & 0.02241 & 0.02241 & 0.02241 & 0.02241\\
$\sin^2\theta_{23}$  & 0.5800  & 0.5800 & 0.5800 & 0.5800 & 0.5800 & 0.5800\\
$\delta_{CP}/\pi$ & 0.556  & 1.391 & 1.20 & 1.586 & 0.320 & 0.893 \\
$\alpha_{21}/\pi$ & 0.811 & 1.015 & 0.997 & 1.623 & 1.363 & 0.927\\
$\alpha_{31}/\pi$ & 0.403  & 1.071 & 0.154 & 1.167 & 0.118 & 1.042\\ \hline
$m_1$/eV& 0.0204 & 0.0162 & 0.0335& 0.0048 & 0.0212 & 0.0063\\
$m_2$/eV & 0.0222 & 0.0183 & 0.0346 & 0.0098 & 0.0229 & 0.0107\\
$m_3$/eV & 0.0542 & 0.0528 & 0.0604 & 0.0505 & 0.0545 & 0.0506\\
$\sum_im_i$/eV & 0.0969 & 0.0872 & 0.1285 & 0.0651 & 0.0987 & 0.0677\\
$|m_{ee}|$/eV & 0.0080 & 0.0061 & 0.0131 & 0.0061 & 0.0136  & 0.00036\\ \hline
$\chi^2_{\mathrm{min}}$                          & $10^{-6}$& $10^{-6}$ & $10^{-6}$ &$10^{-7}$& $10^{-7}$ & $10^{-6}$  \\ \hline\hline
\end{tabular}
\caption{The predictions for the best-fit values of the input parameters and observables in the models $\mathcal{B}_{1,2,3}$ and  $\mathcal{D}_{1,2,3}$ with NO ordering. \label{tab:B1B2B3_D1D2D3} }
\end{table}

\section{\label{sec:conclusion}Conclusion}

In this paper we have provided a comprehensive analysis of lepton masses and mixing in theories with $\Gamma_3\cong A_4$ modular symmetry,
where the single modulus field $\tau$ is the unique source of flavour symmetry breaking, with no flavons allowed, and all masses and Yukawa couplings are modular forms. Similar to previous analyses, we have discussed all the simplest neutrino sectors arising from both the Weinberg operator and the type I seesaw mechanism, with lepton doublets and right-handed neutrinos assumed to be triplets of $A_4$. Unlike previous analyses, we have allowed right-handed charged leptons to transform as all combinations of $\mathbf{1}$, $\mathbf{1}'$ and $\mathbf{1}''$ representations of $A_4$, using the simplest different modular weights to break the degeneracy, leading to ten different charged lepton Yukawa matrices, instead of the usual one.

The above considerations imply ten different Weinberg models, labelled as $\mathcal{A}_{1}$-$\mathcal{A}_{10}$, and thirty different type I seesaw models, labelled as $\mathcal{B}_{1}$-$\mathcal{B}_{10}$, $\mathcal{C}_{1}$-$\mathcal{C}_{10}$, $\mathcal{D}_{1}$-$\mathcal{D}_{10}$, which we have analyzed in detail, in the form of extensive sets of figures and tables. The results of the numerical analysis are summarised in table~\ref{tab:poss_models_numerical_result}, where we see that fourteen
models for both NO and IO can accommodate the data, indicated by ``\ding{52}", where the original model corresponds to the case of
$\mathcal{D}_{10}$ and all the other successful models are new.
Interestingly, most of the successful patterns $\mathcal{B}_{9},\,\mathcal{B}_{10},\,\mathcal{D}_{5} \sim \mathcal{D}_{10}$
(apart from $\mathcal{D}_{5}\sim \mathcal{D}_{6}$)
predict tightly constrained values for the mixing parameters
and large neutrino mass observables $|m_{ee}|$ and $m_{\text{min}}$,
together with approximately maximal Dirac phase. There are also other interesting correlations among the mixing parameters for these models.

The most successful models $\mathcal{B}_{9}$, $\mathcal{B}_{10}$, $\mathcal{D}_{5}\sim\mathcal{D}_{10}$ all contain six real free parameters and two overall mass scales, describing the entire lepton sector (three charged lepton masses, three neutrino masses, three lepton mixing angles and three CP violating phases).
These are the minimal models of $\Gamma_3$ modular-invariant supersymmetry theories allowed by experiment. The results presented here provide new opportunities for $A_4$ modular symmetry model building, including possible extensions to the quark sector.

\subsection*{Acknowledgements}
G.-J.\, D. and X.-G.\, L. acknowledges the support of the National Natural Science Foundation of China under Grant Nos
11522546 and 11835013. S.\,F.\,K. acknowledges the STFC Consolidated Grant ST/L000296/1 and the European Union's Horizon 2020 research and innovation programme under the Marie Sk\l{}odowska-Curie grant agreements Elusives ITN No.\ 674896 and InvisiblesPlus RISE No.\ 690575. G.-J.\, D. and X.-G.\, L. are grateful to Dr. Yang Zhang for his kind help on the \texttt{MultiNest} program.
\providecommand{\href}[2]{#2}\begingroup\raggedright\endgroup

\newpage

\begin{thebibliography}{10}

\bibitem{King:2013eh}
S.~F. King and C.~Luhn, ``{Neutrino Mass and Mixing with Discrete Symmetry},''
  \href{http://dx.doi.org/10.1088/0034-4885/76/5/056201}{{\em Rept. Prog.
  Phys.} {\bfseries 76} (2013) 056201},
\href{http://arxiv.org/abs/1301.1340}{{\ttfamily arXiv:1301.1340 [hep-ph]}}.

\bibitem{King:2017guk}
S.~F. King, ``{Unified Models of Neutrinos, Flavour and CP Violation},''
  \href{http://dx.doi.org/10.1016/j.ppnp.2017.01.003}{{\em Prog. Part. Nucl.
  Phys.} {\bfseries 94} (2017) 217--256},
\href{http://arxiv.org/abs/1701.04413}{{\ttfamily arXiv:1701.04413 [hep-ph]}}.

\bibitem{Koide:2007sr}
Y.~Koide, ``{S(4) flavor symmetry embedded into SU(3) and lepton masses and
  mixing},'' \href{http://dx.doi.org/10.1088/1126-6708/2007/08/086}{{\em JHEP}
  {\bfseries 08} (2007) 086},
\href{http://arxiv.org/abs/0705.2275}{{\ttfamily arXiv:0705.2275 [hep-ph]}}.

\bibitem{Banks:2010zn}
T.~Banks and N.~Seiberg, ``{Symmetries and Strings in Field Theory and
  Gravity},'' \href{http://dx.doi.org/10.1103/PhysRevD.83.084019}{{\em Phys.
  Rev.} {\bfseries D83} (2011) 084019},
\href{http://arxiv.org/abs/1011.5120}{{\ttfamily arXiv:1011.5120 [hep-th]}}.

\bibitem{Wu:2012ria}
Y.-L. Wu, ``{SU(3) Gauge Family Symmetry and Prediction for the Lepton-Flavor
  Mixing and Neutrino Masses with Maximal Spontaneous CP Violation},''
  \href{http://dx.doi.org/10.1016/j.physletb.2012.07.020}{{\em Phys. Lett.}
  {\bfseries B714} (2012) 286--294},
\href{http://arxiv.org/abs/1203.2382}{{\ttfamily arXiv:1203.2382 [hep-ph]}}.

\bibitem{Merle:2011vy}
A.~Merle and R.~Zwicky, ``{Explicit and spontaneous breaking of SU(3) into its
  finite subgroups},'' \href{http://dx.doi.org/10.1007/JHEP02(2012)128}{{\em
  JHEP} {\bfseries 02} (2012) 128},
\href{http://arxiv.org/abs/1110.4891}{{\ttfamily arXiv:1110.4891 [hep-ph]}}.

\bibitem{Rachlin:2017rvm}
B.~L. Rachlin and T.~W. Kephart, ``{Spontaneous Breaking of Gauge Groups to
  Discrete Symmetries},'' \href{http://dx.doi.org/10.1007/JHEP08(2017)110}{{\em
  JHEP} {\bfseries 08} (2017) 110},
\href{http://arxiv.org/abs/1702.08073}{{\ttfamily arXiv:1702.08073 [hep-ph]}}.

\bibitem{Luhn:2011ip}
C.~Luhn, ``{Spontaneous breaking of SU(3) to finite family symmetries: a
  pedestrian's approach},''
  \href{http://dx.doi.org/10.1007/JHEP03(2011)108}{{\em JHEP} {\bfseries 03}
  (2011) 108},
\href{http://arxiv.org/abs/1101.2417}{{\ttfamily arXiv:1101.2417 [hep-ph]}}.

\bibitem{King:2018fke}
S.~F. King and Y.-L. Zhou, ``{Spontaneous breaking of $SO(3)$ to finite family
  symmetries with supersymmetry - an $A_4$ model},''
  \href{http://dx.doi.org/10.1007/JHEP11(2018)173}{{\em JHEP} {\bfseries 11}
  (2018) 173},
\href{http://arxiv.org/abs/1809.10292}{{\ttfamily arXiv:1809.10292 [hep-ph]}}.

\bibitem{Altarelli:2008bg}
G.~Altarelli, F.~Feruglio, and C.~Hagedorn, ``{A SUSY SU(5) Grand Unified Model
  of Tri-Bimaximal Mixing from A$_4$},''
  \href{http://dx.doi.org/10.1088/1126-6708/2008/03/052}{{\em JHEP} {\bfseries
  03} (2008) 052},
\href{http://arxiv.org/abs/0802.0090}{{\ttfamily arXiv:0802.0090 [hep-ph]}}.

\bibitem{Burrows:2009pi}
T.~J. Burrows and S.~F. King, ``{A(4) Family Symmetry from SU(5) SUSY GUTs in
  6d},'' \href{http://dx.doi.org/10.1016/j.nuclphysb.2010.04.002}{{\em Nucl.
  Phys.} {\bfseries B835} (2010) 174--196},
\href{http://arxiv.org/abs/0909.1433}{{\ttfamily arXiv:0909.1433 [hep-ph]}}.

\bibitem{Burrows:2010wz}
T.~J. Burrows and S.~F. King, ``{$A_4$ x SU(5) SUSY GUT of Flavour in 8d},''
  \href{http://dx.doi.org/10.1016/j.nuclphysb.2010.08.018}{{\em Nucl. Phys.}
  {\bfseries B842} (2011) 107--121},
\href{http://arxiv.org/abs/1007.2310}{{\ttfamily arXiv:1007.2310 [hep-ph]}}.

\bibitem{deAnda:2018oik}
F.~J. de~Anda and S.~F. King, ``{An $S_4 \times SU(5)$ SUSY GUT of flavour in
  6d},'' \href{http://dx.doi.org/10.1007/JHEP07(2018)057}{{\em JHEP} {\bfseries
  07} (2018) 057},
\href{http://arxiv.org/abs/1803.04978}{{\ttfamily arXiv:1803.04978 [hep-ph]}}.

\bibitem{Adulpravitchai:2009id}
A.~Adulpravitchai, A.~Blum, and M.~Lindner, ``{Non-Abelian Discrete Flavor
  Symmetries from T**2/Z(N) Orbifolds},''
  \href{http://dx.doi.org/10.1088/1126-6708/2009/07/053}{{\em JHEP} {\bfseries
  07} (2009) 053},
\href{http://arxiv.org/abs/0906.0468}{{\ttfamily arXiv:0906.0468 [hep-ph]}}.

\bibitem{Asaka:2001eh}
T.~Asaka, W.~Buchmuller, and L.~Covi, ``{Gauge unification in
  six-dimensions},''
  \href{http://dx.doi.org/10.1016/S0370-2693(01)01324-7}{{\em Phys. Lett.}
  {\bfseries B523} (2001) 199--204},
\href{http://arxiv.org/abs/hep-ph/0108021}{{\ttfamily arXiv:hep-ph/0108021
  [hep-ph]}}.

\bibitem{Altarelli:2006kg}
G.~Altarelli, F.~Feruglio, and Y.~Lin, ``{Tri-bimaximal neutrino mixing from
  orbifolding},'' \href{http://dx.doi.org/10.1016/j.nuclphysb.2007.03.042}{{\em
  Nucl. Phys.} {\bfseries B775} (2007) 31--44},
\href{http://arxiv.org/abs/hep-ph/0610165}{{\ttfamily arXiv:hep-ph/0610165
  [hep-ph]}}.

\bibitem{Adulpravitchai:2010na}
A.~Adulpravitchai and M.~A. Schmidt, ``{Flavored Orbifold GUT - an SO(10) x S4
  model},'' \href{http://dx.doi.org/10.1007/JHEP01(2011)106}{{\em JHEP}
  {\bfseries 01} (2011) 106},
\href{http://arxiv.org/abs/1001.3172}{{\ttfamily arXiv:1001.3172 [hep-ph]}}.

\bibitem{Kobayashi:2006wq}
T.~Kobayashi, H.~P. Nilles, F.~Ploger, S.~Raby, and M.~Ratz, ``{Stringy origin
  of non-Abelian discrete flavor symmetries},''
  \href{http://dx.doi.org/10.1016/j.nuclphysb.2007.01.018}{{\em Nucl. Phys.}
  {\bfseries B768} (2007) 135--156},
\href{http://arxiv.org/abs/hep-ph/0611020}{{\ttfamily arXiv:hep-ph/0611020
  [hep-ph]}}.

\bibitem{deAnda:2018yfp}
F.~J. de~Anda and S.~F. King, ``{$SU(3) \times SO(10)$ in 6d},''
  \href{http://dx.doi.org/10.1007/JHEP10(2018)128}{{\em JHEP} {\bfseries 10}
  (2018) 128},
\href{http://arxiv.org/abs/1807.07078}{{\ttfamily arXiv:1807.07078 [hep-ph]}}.

\bibitem{Kobayashi:2018rad}
T.~Kobayashi, S.~Nagamoto, S.~Takada, S.~Tamba, and T.~H. Tatsuishi, ``{Modular
  symmetry and non-Abelian discrete flavor symmetries in string
  compactification},'' \href{http://dx.doi.org/10.1103/PhysRevD.97.116002}{{\em
  Phys. Rev.} {\bfseries D97} no.~11, (2018) 116002},
\href{http://arxiv.org/abs/1804.06644}{{\ttfamily arXiv:1804.06644 [hep-th]}}.

\bibitem{Baur:2019kwi}
A.~Baur, H.~P. Nilles, A.~Trautner, and P.~K.~S. Vaudrevange, ``{Unification of
  Flavor, CP, and Modular Symmetries},''
\href{http://arxiv.org/abs/1901.03251}{{\ttfamily arXiv:1901.03251 [hep-th]}}.

\bibitem{Giveon:1988tt}
A.~Giveon, E.~Rabinovici, and G.~Veneziano, ``{Duality in String Background
  Space},''
\href{http://dx.doi.org/10.1016/0550-3213(89)90489-6}{{\em Nucl. Phys.}
  {\bfseries B322} (1989) 167--184}.

\bibitem{Altarelli:2005yx}
G.~Altarelli and F.~Feruglio, ``{Tri-bimaximal neutrino mixing, A(4) and the
  modular symmetry},''
  \href{http://dx.doi.org/10.1016/j.nuclphysb.2006.02.015}{{\em Nucl. Phys.}
  {\bfseries B741} (2006) 215--235},
\href{http://arxiv.org/abs/hep-ph/0512103}{{\ttfamily arXiv:hep-ph/0512103
  [hep-ph]}}.

\bibitem{deAdelhartToorop:2011re}
R.~de~Adelhart~Toorop, F.~Feruglio, and C.~Hagedorn, ``{Finite Modular Groups
  and Lepton Mixing},''
  \href{http://dx.doi.org/10.1016/j.nuclphysb.2012.01.017}{{\em Nucl. Phys.}
  {\bfseries B858} (2012) 437--467},
\href{http://arxiv.org/abs/1112.1340}{{\ttfamily arXiv:1112.1340 [hep-ph]}}.

\bibitem{Feruglio:2017spp}
F.~Feruglio, \href{http://dx.doi.org/10.1142/9789813238053_0012}{``{Are
  neutrino masses modular forms?},''} in {\em From My Vast Repertoire ...:
  Guido Altarelli's Legacy}, A.~Levy, S.~Forte, and G.~Ridolfi, eds.,
  pp.~227--266.
\newblock 2019.
\newblock
\href{http://arxiv.org/abs/1706.08749}{{\ttfamily arXiv:1706.08749 [hep-ph]}}.
\newblock

\bibitem{Criado:2018thu}
J.~C. Criado and F.~Feruglio, ``{Modular Invariance Faces Precision Neutrino
  Data},''
\href{http://arxiv.org/abs/1807.01125}{{\ttfamily arXiv:1807.01125 [hep-ph]}}.

\bibitem{Kobayashi:2018vbk}
T.~Kobayashi, K.~Tanaka, and T.~H. Tatsuishi, ``{Neutrino mixing from finite
  modular groups},'' \href{http://dx.doi.org/10.1103/PhysRevD.98.016004}{{\em
  Phys. Rev.} {\bfseries D98} no.~1, (2018) 016004},
\href{http://arxiv.org/abs/1803.10391}{{\ttfamily arXiv:1803.10391 [hep-ph]}}.

\bibitem{Kobayashi:2018wkl}
T.~Kobayashi, Y.~Shimizu, K.~Takagi, M.~Tanimoto, T.~H. Tatsuishi, and
  H.~Uchida, ``{Finite modular subgroups for fermion mass matrices and
  baryon/lepton number violation},''
\href{http://arxiv.org/abs/1812.11072}{{\ttfamily arXiv:1812.11072 [hep-ph]}}.

\bibitem{Kobayashi:2019rzp}
T.~Kobayashi, Y.~Shimizu, K.~Takagi, M.~Tanimoto, and T.~H. Tatsuishi,
  ``{Modular $S_3$ invariant flavor model in SU(5) GUT},''
\href{http://arxiv.org/abs/1906.10341}{{\ttfamily arXiv:1906.10341 [hep-ph]}}.

\bibitem{Okada:2019xqk}
H.~Okada and Y.~Orikasa, ``{A modular $S_3$ symmetric radiative seesaw
  model},''
\href{http://arxiv.org/abs/1907.04716}{{\ttfamily arXiv:1907.04716 [hep-ph]}}.

\bibitem{Kobayashi:2018scp}
T.~Kobayashi, N.~Omoto, Y.~Shimizu, K.~Takagi, M.~Tanimoto, and T.~H.
  Tatsuishi, ``{Modular $A_4$ invariance and neutrino mixing},''
\href{http://arxiv.org/abs/1808.03012}{{\ttfamily arXiv:1808.03012 [hep-ph]}}.

\bibitem{Okada:2018yrn}
H.~Okada and M.~Tanimoto, ``{CP violation of quarks in $A_4$ modular
  invariance},''
\href{http://arxiv.org/abs/1812.09677}{{\ttfamily arXiv:1812.09677 [hep-ph]}}.

\bibitem{Novichkov:2018yse}
P.~P. Novichkov, S.~T. Petcov, and M.~Tanimoto, ``{Trimaximal Neutrino Mixing
  from Modular A$_4$ Invariance with Residual Symmetries},''
\href{http://arxiv.org/abs/1812.11289}{{\ttfamily arXiv:1812.11289 [hep-ph]}}.

\bibitem{Nomura:2019yft}
T.~Nomura and H.~Okada, ``{A two loop induced neutrino mass model with modular
  $A_4$ symmetry},''
\href{http://arxiv.org/abs/1906.03927}{{\ttfamily arXiv:1906.03927 [hep-ph]}}.

\bibitem{Penedo:2018nmg}
J.~T. Penedo and S.~T. Petcov, ``{Lepton Masses and Mixing from Modular $S_4$
  Symmetry},'' \href{http://dx.doi.org/10.1016/j.nuclphysb.2018.12.016}{{\em
  Nucl. Phys.} {\bfseries B939} (2019) 292--307},
\href{http://arxiv.org/abs/1806.11040}{{\ttfamily arXiv:1806.11040 [hep-ph]}}.

\bibitem{Novichkov:2018ovf}
P.~P. Novichkov, J.~T. Penedo, S.~T. Petcov, and A.~V. Titov, ``{Modular $S_4$
  Models of Lepton Masses and Mixing},''
\href{http://arxiv.org/abs/1811.04933}{{\ttfamily arXiv:1811.04933 [hep-ph]}}.

\bibitem{Kobayashi:2019mna}
T.~Kobayashi, Y.~Shimizu, K.~Takagi, M.~Tanimoto, and T.~H. Tatsuishi, ``{New
  $A_4$ lepton flavor model from $S_4$ modular symmetry},''
\href{http://arxiv.org/abs/1907.09141}{{\ttfamily arXiv:1907.09141 [hep-ph]}}.

\bibitem{Novichkov:2018nkm}
P.~P. Novichkov, J.~T. Penedo, S.~T. Petcov, and A.~V. Titov, ``{Modular $A_5$
  Symmetry for Flavour Model Building},''
\href{http://arxiv.org/abs/1812.02158}{{\ttfamily arXiv:1812.02158 [hep-ph]}}.

\bibitem{Ding:2019xna}
G.-J. Ding, S.~F. King, and X.-G. Liu, ``{Neutrino Mass and Mixing with $A_5$
  Modular Symmetry},''
\href{http://arxiv.org/abs/1903.12588}{{\ttfamily arXiv:1903.12588 [hep-ph]}}.

\bibitem{deAnda:2018ecu}
F.~J. de~Anda, S.~F. King, and E.~Perdomo, ``{$SU(5)$ Grand Unified Theory with
  $A_4$ Modular Symmetry},''
\href{http://arxiv.org/abs/1812.05620}{{\ttfamily arXiv:1812.05620 [hep-ph]}}.

\bibitem{deMedeirosVarzielas:2019cyj}
I.~de~Medeiros~Varzielas, S.~F. King, and Y.-L. Zhou, ``{Multiple modular
  symmetries as the origin of flavour},''
\href{http://arxiv.org/abs/1906.02208}{{\ttfamily arXiv:1906.02208 [hep-ph]}}.

\bibitem{Novichkov:2019sqv}
P.~P. Novichkov, J.~T. Penedo, S.~T. Petcov, and A.~V. Titov, ``{Generalised CP
  Symmetry in Modular-Invariant Models of Flavour},''
\href{http://arxiv.org/abs/1905.11970}{{\ttfamily arXiv:1905.11970 [hep-ph]}}.

\bibitem{Liu:2019khw}
X.-G. Liu and G.-J. Ding, ``{Neutrino Masses and Mixing from Double Covering of
  Finite Modular Groups},''
\href{http://arxiv.org/abs/1907.01488}{{\ttfamily arXiv:1907.01488 [hep-ph]}}.

\bibitem{Bruinier2008The}
J.~H. Bruinier, G.~V.~D. Geer, G.~Harder, and D.~Zagier, {\em The 1-2-3 of
  Modular Forms}.
\newblock Universitext. Springer Berlin Heidelberg, 2008.

\bibitem{diamond2005first}
F.~Diamond and J.~M. Shurman, {\em A first course in modular forms}, vol.~228
  of {\em Graduate Texts in Mathematics}.
\newblock Springer, 2005.

\bibitem{Gunning1962}
R.~C. Gunning, {\em Lectures on Modular Forms}.
\newblock Princeton, New Jersey USA, Princeton University Press, 1962.

\bibitem{Esteban:2018azc}
I.~Esteban, M.~C. Gonzalez-Garcia, A.~Hernandez-Cabezudo, M.~Maltoni, and
  T.~Schwetz, ``{Global analysis of three-flavour neutrino oscillations:
  synergies and tensions in the determination of $\theta_{23}, \delta_{CP}$,
  and the mass ordering},''
\href{http://arxiv.org/abs/1811.05487}{{\ttfamily arXiv:1811.05487 [hep-ph]}}.

\bibitem{Feruglio:2014jla}
F.~Feruglio, K.~M. Patel, and D.~Vicino, ``{Order and Anarchy hand in hand in
  5D SO(10)},'' \href{http://dx.doi.org/10.1007/JHEP09(2014)095}{{\em JHEP}
  {\bfseries 09} (2014) 095},
\href{http://arxiv.org/abs/1407.2913}{{\ttfamily arXiv:1407.2913 [hep-ph]}}.

\bibitem{Ross:2007az}
G.~Ross and M.~Serna, ``{Unification and fermion mass structure},''
  \href{http://dx.doi.org/10.1016/j.physletb.2008.05.014}{{\em Phys. Lett.}
  {\bfseries B664} (2008) 97--102},
\href{http://arxiv.org/abs/0704.1248}{{\ttfamily arXiv:0704.1248 [hep-ph]}}.

\bibitem{Feroz:2007kg}
F.~Feroz and M.~P. Hobson, ``{Multimodal nested sampling: an efficient and
  robust alternative to MCMC methods for astronomical data analysis},''
  \href{http://dx.doi.org/10.1111/j.1365-2966.2007.12353.x}{{\em Mon. Not. Roy.
  Astron. Soc.} {\bfseries 384} (2008) 449},
\href{http://arxiv.org/abs/0704.3704}{{\ttfamily arXiv:0704.3704 [astro-ph]}}.

\bibitem{Feroz:2008xx}
F.~Feroz, M.~P. Hobson, and M.~Bridges, ``{MultiNest: an efficient and robust
  Bayesian inference tool for cosmology and particle physics},''
  \href{http://dx.doi.org/10.1111/j.1365-2966.2009.14548.x}{{\em Mon. Not. Roy.
  Astron. Soc.} {\bfseries 398} (2009) 1601--1614},
\href{http://arxiv.org/abs/0809.3437}{{\ttfamily arXiv:0809.3437 [astro-ph]}}.

\bibitem{Aghanim:2018eyx}
{\bfseries Planck} Collaboration, N.~Aghanim {\em et~al.}, ``{Planck 2018
  results. VI. Cosmological parameters},''
\href{http://arxiv.org/abs/1807.06209}{{\ttfamily arXiv:1807.06209
  [astro-ph.CO]}}.

\bibitem{Tanabashi:2018oca}
{\bfseries Particle Data Group} Collaboration, M.~Tanabashi {\em et~al.},
  ``{Review of Particle Physics},''
\href{http://dx.doi.org/10.1103/PhysRevD.98.030001}{{\em Phys. Rev.} {\bfseries
  D98} no.~3, (2018) 030001}.

\end{thebibliography}
\end{document}